\documentclass[fleqn,usenatbib]{mnras}
\usepackage{newtxtext,newtxmath}

\usepackage[utf8]{inputenc}
\usepackage[T1]{fontenc}
\usepackage{ae,aecompl}
\usepackage{graphicx}	
\usepackage{amsmath}	
\usepackage{amssymb}	
\usepackage[normalem]{ulem}
\usepackage{multicol}
\usepackage{pdflscape}
\usepackage{lineno}

\newcommand{\jaavso}{JAAVSO}

\title[Dust Declines in RCB Stars]{A Comprehensive Study of the Dust Declines in R Coronae Borealis Stars}
\author[C. L. Crawford et al.]{
Courtney L. Crawford$^{1}$\thanks{Email: courtney.crawford@sydney.edu.au},
Jamie Soon$^{2}$,
Geoffrey C. Clayton$^{3,4}$,
Patrick Tisserand$^{5}$,
\newauthor Timothy R. Bedding$^{1}$,
Caleb J. Clark$^{1}$,
Chung-Uk Lee$^{6}$
\\
$^{1}$ Sydney Institute for Astronomy (SIfA), School of Physics, University of Sydney, NSW 2006, Australia\\
$^{2}$ Research School of Astronomy and Astrophysics, Australian National University, Cotter Rd, Weston Creek ACT 2611, Australia\\
$^{3}$ Department of Physics \& Astronomy, Louisiana State University, Baton Rouge, LA 70803, USA\\
$^{4}$ Space Science Institute, 4765 Walnut St, Suite B, Boulder, CO 80301, USA\\
$^{5}$ Sorbonne Universit\'es, UPMC Univ. Paris 6 et CNRS, UMR 7095, Institut d’Astrophysique de Paris, IAP, 75014 Paris, France \\
$^{6}$ Korea Astronomy and Space Science Institute, Daejeon 34055, Republic of Korea\\
}
\date{Accepted 2025 February 3. Received 2025 February 2; in original form 2024 December 20}
\pubyear{2024}
\begin{document}
\label{firstpage}
\pagerange{\pageref{firstpage}--\pageref{lastpage}}
\maketitle

\begin{abstract}
The R Coronae Borealis (RCB) variables are rare, hydrogen-deficient, carbon-rich supergiants known for large, erratic declines in brightness due to dust formation. Recently, the number of known RCB stars in the Milky Way and Magellanic Clouds has increased from $\sim$30 to 162. We use all-sky and targeted photometric surveys to create the longest possible light curves for all known RCB stars and systematically study their declines. Our study, the largest of its kind, includes measurements of decline activity levels, morphologies, and periodicities for nearly all RCB stars. We confirm previous predictions that cool RCB stars exhibit more declines than warm RCBs, supporting a relationship between dust formation and condensation temperatures. We also find evidence for two distinct dust production mechanisms. R CrB and SU Tau show decline onsets consistent with a Poisson process, suggesting their dust production is driven by stochastic processes, such as convection. In contrast, RY Sgr’s declines correlate with its pulsation period, suggesting that its dust production is driven by pulsationally-induced shocks. Finally, we show that the dust properties of the related class of DY~Per variables differ from those of the RCB stars, suggesting differences in their evolutionary status.
\end{abstract}

\begin{keywords}
stars: chemically peculiar - stars: carbon - stars: variables: general - stars: activity
\end{keywords}


\section{Introduction} 
In 1784, R Coronae Borealis was one of the first variable stars to be discovered when it seemed to disappear from the sky \citep{1797RSPT...87..133P}. Its discovery, and until recently the discoveries of most R Coronae Borealis (RCB) variables, were made due to their large, irregular declines in brightness.  
The RCBs are a small class of carbon-rich, hydrogen-deficient
supergiants that produce dust, which create spectacular declines in brightness \citep{1996PASP..108..225C, 2012JAVSO..40..539C}. There is an even smaller set of stars which are spectroscopically similar to the RCB stars, but do not produce dust \citep{2022A&A...667A..83T}. These are the dustless hydrogen-deficient carbon (dLHdC) stars. These two classes of variables form an umbrella class of stars known as the hydrogen-deficient carbon (HdC) stars.

There are two types of brightness variations seen in RCB stars--- small amplitude ($\leq0.3$ mag) semi-regular variations typically attributed to stellar pulsations, and large amplitude dips (up to 9 mag) in brightness which we refer to as ``declines''. The declines have been known for nearly a century to be the product of thick carbon dust production \citep{1935AN....254..151L,1939ApJ....90..294O}, however the exact dust formation mechanism is still unknown \citep{1996A&A...313..217W}. While regular patterns in the RCB declines would favour mechanisms such as orbiting dust clouds, stellar rotation of starspots, or some kind of binary interaction \citep[][and references therein]{1996PASP..108..225C}, the RCB star declines occur at completely irregular intervals \citep{1993PASP..105..832C}. The only regularity discovered is a possible relationship between the phase of pulsation and the onset of declines for a few RCB stars \citep{2007MNRAS.375..301C,2018JAVSO..46..127P}. 

There have been two systematic studies of RCB declines. \citet{1996AcA....46..325J} studied all available light curves for about 20 bright RCB stars and found a wide range in the frequency of declines from star to star. She also found that stars with larger H abundances had more frequent declines. \citet{2024MNRAS.527.9274S} recently studied the declines for 10 bright RCB stars using light curves from the American Association of Variable Star Observers (AAVSO) and archival plates from the Harvard College Observatory, including those from the Digital Access to a Sky Century at Harvard (DASCH) project \citep{2012IAUS..285...29G}. He focused on whether RCB stars were experiencing long-term ($\sim$century long timescale) brightenings which would be consistent with evolution leftward in the Hertzsprung-Russell Diagram (HRD) towards the Extreme Helium (EHe) stars. He did not find this characteristic brightening in any of the studied RCBs, though it has been measured for the hottest RCBs previously \citep{2002AJ....123.3387D,2016MNRAS.460.1233S}. He additionally explored some of the decline statistics for these 10 stars such as decline frequencies, lengths, and depths, and additionally showed that the decline recovery phases of isolated declines have similar morphologies. 

Although there have been few systematic studies of RCB declines, many facts are known about these declines. The RCB stars are understood to show a wide range of decline activity. For example, there are a few known RCB stars which have famously spent significant amounts of time in decline, such as V854 Cen, the third brightest RCB in the sky at 7th magnitude, but undiscovered until the 1980s \citep{1986IAUC.4233....3M}. It was generally fainter than 13th magnitude and no brighter than 10th from 1913 to 1952, so it does not appear in the HD, CPD or SAO catalogs \citep{1986IAUC.4241....2W,1986IAUC.4245....2M}. A less extreme case is the prototype R CrB itself, which in 2007 began a decline that lasted for roughly 12 years. In addition to the known long-lived declines, there are RCB stars which decline more than once per year, such as UW Cen. However, there are also known RCB stars which exhibit very few declines, such as XX Cam and Y Mus \citep{1928BHarO.861...11P,1948ApJ...107..413Y,1980JApA....1...71K}. However, these seemingly inactive stars still have bright dust shells in the IR, indicating that dust is being actively produced, just not in our line of sight \citep{1997MNRAS.285..317F}. This behaviour implies that the formation of RCB dust is not symmetrical around the star. Polarimetric observations of R CrB itself suggest that its dust lies in an obscuring equatorial torus and polar dust lobes \citep{1997ApJ...476..870C}. 

In recent years, IR color cuts and spectroscopic followup have been used in the Milky Way and Magellanic Clouds to discover significantly more RCB stars (now totalling 162 known stars), especially those with cooler surface temperatures \citep{2012A&A...539A..51T,2013A&A...551A..77T,2020A&A...635A..14T,2022A&A...667A..83T,2021ApJ...910..132K,2024PASP..136h4201K}.
However, there have not been significant improvements in the estimations of T$_{\rm eff}$ or abundances for the RCB stars for over 20 years \citep{2000A&A...353..287A}, since hydrogen-deficient, carbon-rich, metal-poor atmospheric models are not commonly available. However, low/medium-resolution spectra used for spectroscopic follow-up has enabled the creation of a spectral classification system for the HdC stars, which can be used as a proxy for T$_{\rm eff}$ estimation in RCBs \citep{2023MNRAS.521.1674C}. 

The DY Persei-like variables are another class of star that show similar variability to the RCB stars, though the extent of their connection to the RCB variables themselves is currently not understood \citep{2009A&A...501..985T,Crawford2024_dlhdcmodels}. They are sometimes included under the HdC umbrella, although their hydrogen-deficiency remains unconfirmed \citep{2018ApJ...854..140B,2023ApJ...948...15G}. The DY Per variables are known to show similar dust declines to the RCB stars, although DY Per declines are both shallower and more symmetrical than the RCB declines \citep{2001ApJ...554..298A}. Given that these stars are potentially cooler members of the HdC class, we included them in portions of our analysis when relevant\footnote{For clarity: we include the DY Per stars as their own cool HdC class in Section~\ref{sec:decline_rate}, and DY Per itself is included in Section~\ref{subsec:slopes}. Otherwise, the DY Pers are excluded from the total sample.} in the hope of shedding light on the potential connections and differences between these two relatively unstudied classes of variables. 

In this paper, we take advantage of the new, larger sample of 162 known RCB stars, as well as the plethora of light curves available from various all-sky and targeted photometric surveys. We investigate the relationship of the decline characteristics---such as frequency---to stellar characteristics such as temperature and abundances, with the goal of a more complete understanding of the RCB dust formation mechanisms and dust behaviour. 
In Section~\ref{sec:data} we describe our data collection and preprocessing. In Section~\ref{sec:detection} we explain our methods for identifying decline onset and ending times. In Section~\ref{sec:decline_rate} we show our main decline activity indicators and how they vary with the temperature of the stars. In Section~\ref{sec:decline_morph} we discuss the three aspects of the decline morphologies, namely the depths, the onset slopes, and the recovery shapes. In Section~\ref{sec:decline_onsets} we show the waiting time distributions for RCB onsets and compare them to existing models. Finally, we conclude and summarize in Section~\ref{sec:conclusions}.

\section{Data}
\label{sec:data}

As the study of RCB declines require the longest possible baselines, we combined light curves from as many different photometric sources as possible. The brightest RCB stars are both historically relevant and have been popular targets for amateur observations, so the best source for these stars is the AAVSO (American Association of Variable Star Observers; \citealt{Kloppenborg23}). However, collecting long baseline light curves for the fainter RCB stars is much more challenging. The recent uptick in the number of wide-field all-sky surveys means that there is currently a large amount of data on the RCB stars. For most RCB stars, the most complete light curve will contain data from multiple different sources, with different passbands, saturation and plate limits, and duty cycles. For our study, we used a collection of many different photometric sources: AAVSO \citep{Kloppenborg23}, ASAS-3 \citep{Pojmanski1997_ASAS}, ASAS-SN \citep{2018MNRAS.477.3145J,2019MNRAS.486.1907J}, OGLE \citep{2008AcA....58..187U}, EROS \citep{1997NIMPA.387..286B}, KMTNET \citep{Lee2014_kmtnet1,Kim2016_kmtnet2}, ATLAS \citep{Tonry2018_atlas}, ZTF \citep{Masci2019_ztf}, and Gaia DR3 \citep{GaiaDR3_release}. We do not use Kepler, K2, or TESS because the Kepler main mission does not have any RCB stars in its footprint, and both K2 and TESS's typical observation time for each RCB star is too short to be used for decline studies. 

AAVSO data are mostly collected by amateur astronomers, and the longest baseline data contains visual (by-eye) measurements. This comes with its own set of challenges, as the brightness data are by nature discretized to 0.1 mag intervals, and there may be systematic errors associated with multiple observers. Therefore, the usage of AAVSO data necessitates a bit of preprocessing. For this, we first removed all AAVSO observations from observers who contributed fewer than 5 observations to the light curve. We then binned the light curves to 5 day intervals. We did not include measurements which were reported as upper limits. 

There are a number of additional challenges that should be kept in mind. Firstly, nearly all RCB star light curves will have a gap each year when the star is not observable and our study will miss any declines that occur during this gap. This is partially mitigated by the use of multiple different light curve sources. Secondly, each photometric survey will have a different passband, and therefore the light curve will differ slightly in each passband, especially in the depths of the declines. This change in colour contains information on the dust, which we plan to analyse in a future work. However, there are also great advantages to combining multiple data sets. The most obvious is the vast expansion of the total observation time. However, the different saturation limits and plate limits means that some surveys are better at observing RCB stars in different phases of their declines. 


\begin{figure*}
    \centering
    \includegraphics[width=\textwidth]{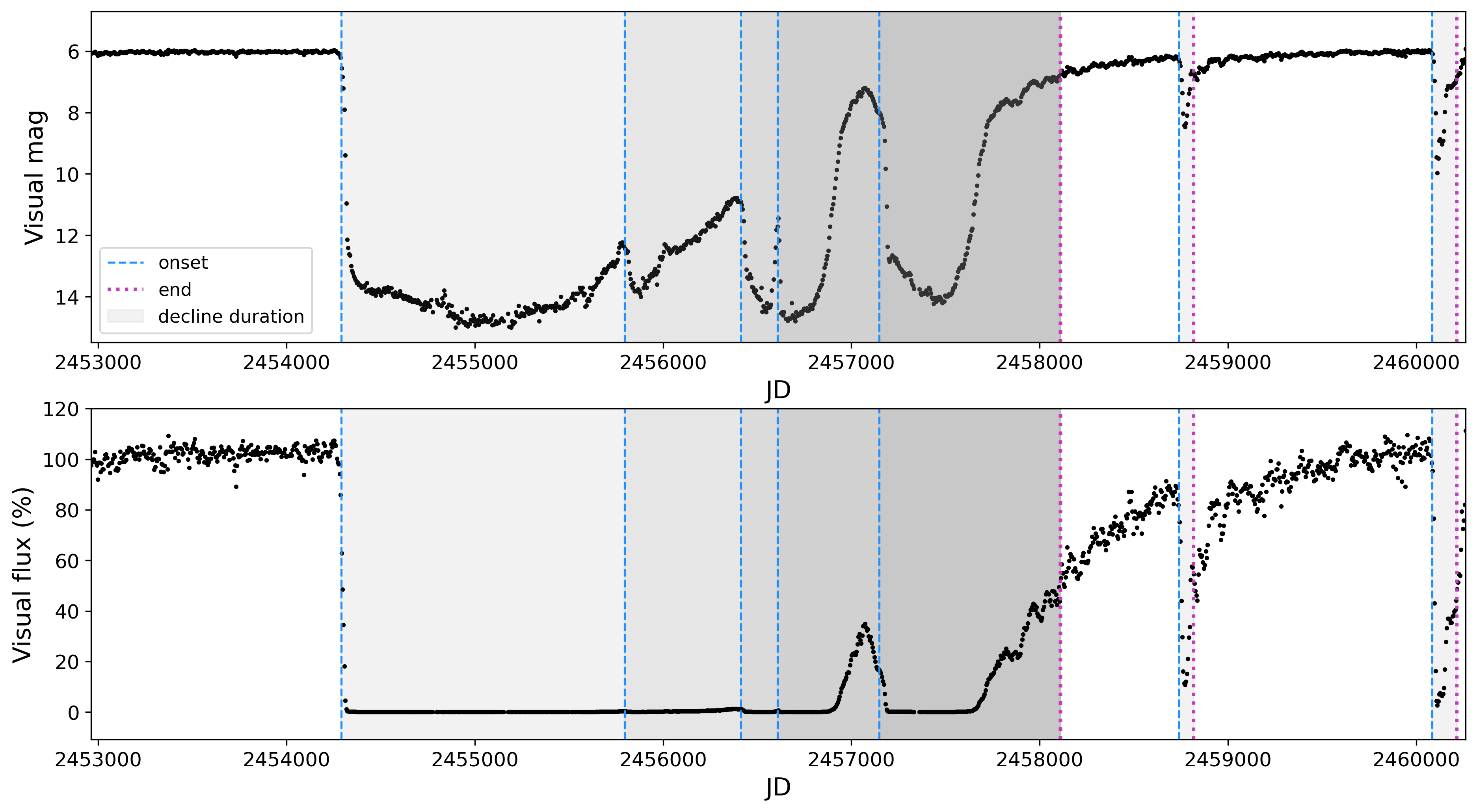}
    \caption{Twenty year AAVSO visual light curve of R CrB ending on 14 November, 2023. The upper panel shows the light curve in units of magnitudes, whereas the lower panel shows the light curve as a percentage of total stellar flux at maximum light. The blue dashed lines indicate our detected decline onsets, and the pink dotted lines denote the adopted end time for each decline. The shaded regions indicate the timespan of individual declines, such that the darker shaded regions indicate how many nested declines are within one large decline event.}
    \label{fig:rcrb-light curve}
\end{figure*}

\section{Decline Detection}
\label{sec:detection}

We combined all available data for each RCB star into one light curve. For each individual passband light curve, we first used a non-uniform Savitzky-Golay filter to remove outliers. For each star, we located the onset and end of each decline in the full-bandpass light curve by visual inspection. We defined a {\em decline} as a dip in brightness of $>$ 1 mag\footnote{This depth should be different for each bandpass. However, the bandpasses used in this analysis are similar enough that the 1 mag limit is sufficient for all data, and should not affect the identification of declines in any individual light curve.} from the star's maximum brightness. We defined the {\em onset of a decline} as the last time where the flux was observed at maximum light before the decline. We defined the {\em end of a decline} as the time when the flux returned to within 1 mag of its maximum light. We note that defining the end of a decline end is much more ambiguous than its onset because the recovery phase of any decline asymptotes towards maximum light \citep{2024MNRAS.527.9274S}. 

In some cases, RCB stars exhibit what we call {\em nested declines}, with a partial recovery in brightness after the onset of a decline, but then followed by another sharp decline. In this case, we defined each drop in flux as a separate decline event, with the second decline nested within the first decline. We defined these two declines as having the same end time, but different onset times. The onset for a nested decline was defined slightly differently, since it does not start from maximum light. Instead, we defined the onset time for a nested decline to be the time of local maximum in brightness immediately prior to the decline. A nested decline need not have any minimum rise in brightness (i.e. a nested decline does not need to have depth $>$1 magnitude to be counted, since some nested declines are diluted by the larger decline event).

We show an example of decline identifications in Figure~\ref{fig:rcrb-light curve} using the last twenty years of the R CrB AAVSO light curve, which includes its historic 2007 decline. The upper panel shows the light curve in magnitude space, which is how they are typically viewed. In this representation, it is clear where each individual decline event happens, including the multiple nested declines which occur inside the long-term decline event. In the lower panel, for illustrative purposes, we show the light curve as a percentage of the maximum light stellar flux. Here it is plain to see that the behaviour of the light at the decline minima (or in other words, small nested decline events) is a very subtle effect in the actual brightness of the star. We remind the reader that a change in magnitude of 1 magnitude implies a $\sim$60\% decrease in flux. In the flux-percentage representation, it is also easy to see on the right-hand side the difficulty of measuring the end of a decline, as the brightness has not truly returned to 100\% flux before its next decline. The flux graph also makes it clear that the declines involve almost complete blockage of light from the star.

Since our data are ground-based and often have significant gaps between data points, we took care to ensure proper error estimation in the decline onset and end times. When a decline onset or end was implied to occur during a gap in the data, we quote our best estimate of the decline onset/end based on the surrounding data points. Each detected onset and end time is given an associated error estimate both on the right side (increasing age) and the left side (decreasing age) based on the average age difference of the nearest 5 data points on the right and left sides, respectively. Therefore, decline times quoted in a gap will have large errors that reflect the effects of such a gap. We additionally provide quality flags for the onset and end times, which indicate whether this detection occurred within a significant ($>$1 year) gap.

We measured a total of 1536 declines over the 162 RCB stars in our sample. Of these, 758 declines (49\%) were fully isolated (i.e. not nested). Among the rest, 497 declines (32\% of the total) were nested within a larger decline. Decline events with nested declines have an average of 2.75 total declines per event. Only $\sim$6\% of isolated declines were longer than 2 years. 

\subsection{Why not Automated?}
\label{subsec:automation_issues}

The use of manual decline detection certainly introduces a subjective element into our analysis. However, due to the complexity of our data, automated methods are currently not feasible for this work. The most important problem in introducing automation to this work is how to properly interpolate and smooth over large gaps in the data. The majority of edge-detection methods currently available in the literature generally rely on having evenly sampled data, and often require the data to be smooth. Therefore, automated techniques for this problem will have to make a decision on how to in-paint during the gaps in the data. For some stars, these gaps are short and can be easily interpolated over. However, for some stars the gaps last for 5 years or more. The stars studied here are far too unpredictable to say anything during those gaps in the data. The usage of multiple photometric sources helps significantly with filling in the gaps in certain datasets, but in order to then automate the decline detection, one must then attempt to correct multiple different light curves into one photometric system, which would need to take into account the differing depths of the declines in each photometric band. This type of analysis is currently outside the scope of this work.

\section{Decline Activity Statistics}
\label{sec:decline_rate}

\begin{figure}
    \centering
    \includegraphics[width=\columnwidth]{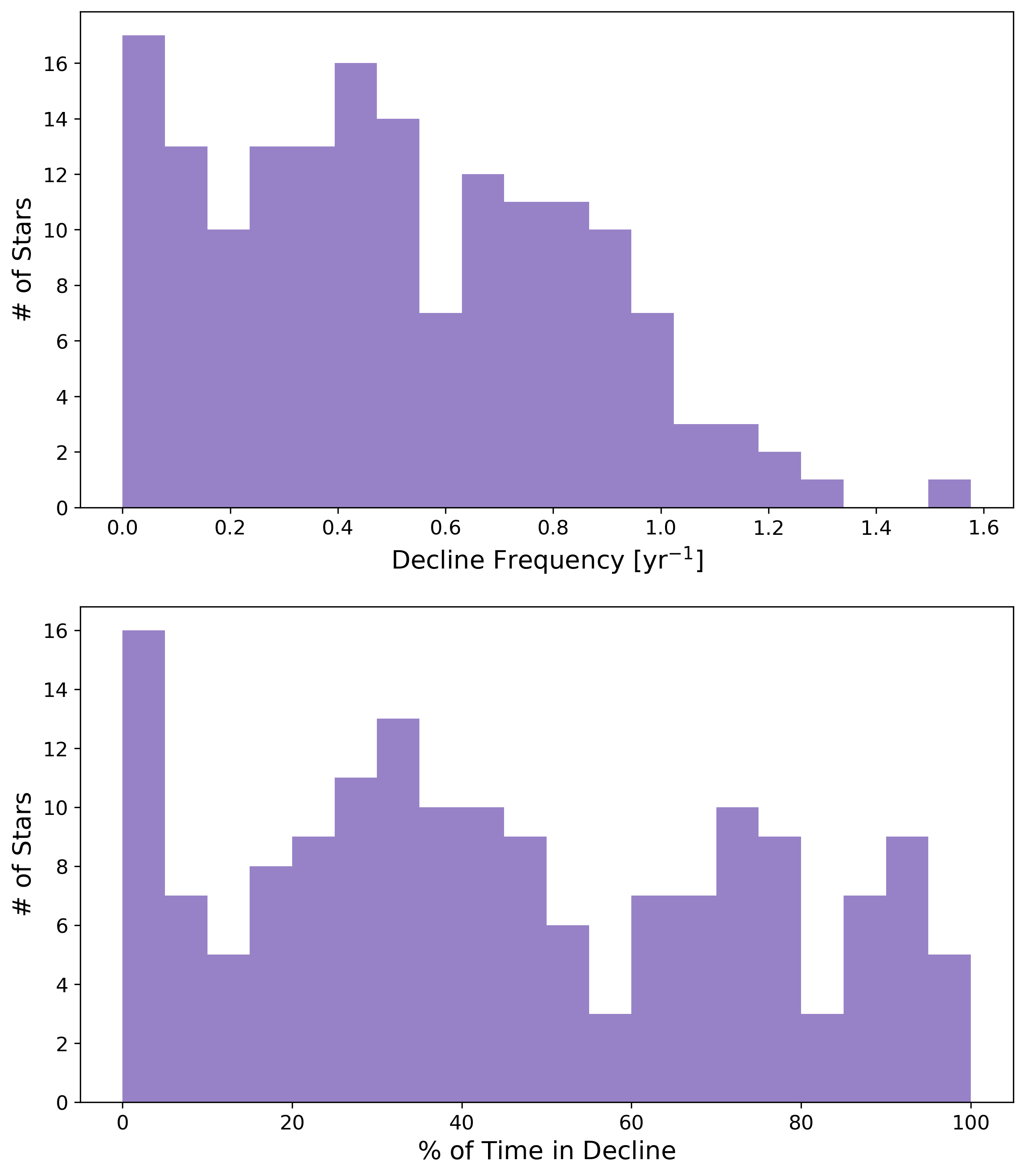}
\caption{Histograms of the decline activity statistics for all 162 measured RCB stars. The upper panel shows the frequency of declines per year and the lower panel shows the percentage of time spent in decline.}
\label{fig:stats_histograms}
\end{figure}

For each RCB star, we calculated two decline activity statistics: the decline frequency and the percentage of time in decline. The former is defined as the number of declines detected divided by the baseline of observations, which we calculated as the time elapsed from the first observation to the last, minus the summed length of all gaps longer than one year. Decline frequency is equivalent to the average number of declines per year. The percentage of time in decline was measured as the total time spent in decline (minus gaps longer than one year) divided by the time baseline of observations. We show the distributions of decline frequency and percent of time in decline for all RCB stars in Figure~\ref{fig:stats_histograms}. From these, it is clear that RCBs span a very large range of decline activity levels. 

\begin{figure}
    \centering
    \includegraphics[width=\columnwidth]{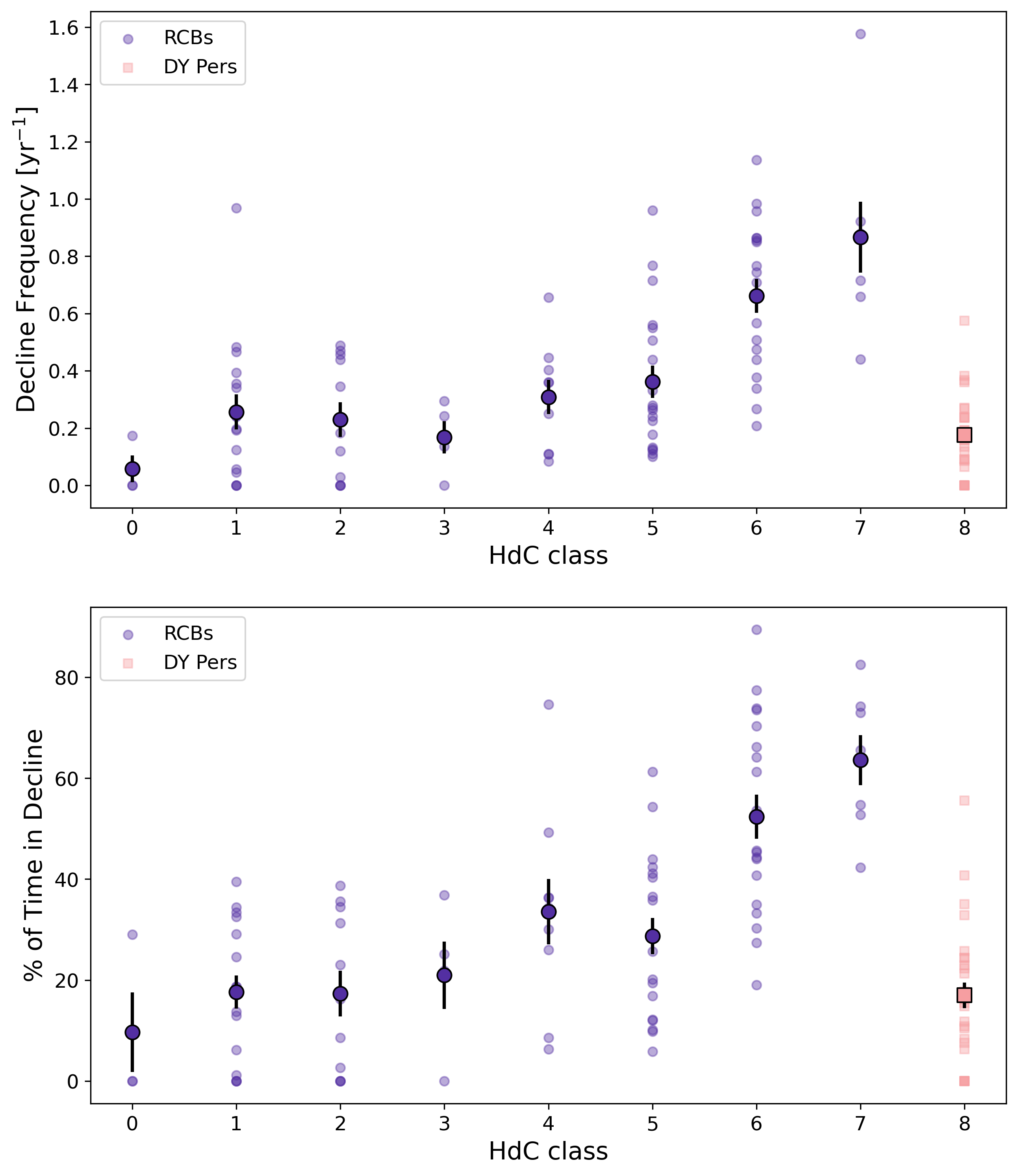}
    \caption{Decline statistics versus the HdC temperature class (0 is warmest, 7 is coolest, and 8 denotes the DY Per stars). The upper panel shows the decline frequency and the lower panel shows the percent of time spent in decline for the RCB stars (purple circles), and the DY Per stars (peach squares). The larger filled points denote the mean for each HdC class, and the error bars denote the standard error of the mean.}
    \label{fig:activity_vs_class}
\end{figure}

Recently, \citet{2024A&A...684A.130T} investigated the dust production rates in RCB stars by taking the ratio of mean Gaia $G$-band flux to its standard error, which is a good indicator for large brightness variations during the $\sim$3 yrs covered by Gaia DR3. Using these metrics, they found suggestions that the cooler RCB stars (as measured by their spectral type from \citealt{2023MNRAS.521.1674C}) have declines more often than the warmer RCB stars. In Figure~\ref{fig:activity_vs_class} we show our two decline activity statistics versus the HdC class from \citet{2023MNRAS.521.1674C} for the 115 stars for which this temperature classification is available. From our data, which are based on much longer time series than the Gaia DR3 used in \citet{2024A&A...684A.130T}, it is clear that the cooler RCB stars not only exhibit more frequent declines, but also spend more time in decline than their warmer counterparts. In fact, many cooler RCB stars are in prolonged declines and have never been observed at maximum light \citep{2024PASP..136h4201K}. For the warmer stars, the decline frequency is roughly constant, but then increases steadily from Class 3 to the cooler classes. 

The percent of time spent in decline shows a similar trend to that of the decline frequency, but even more pronounced, as shown in the lower panel of Fig.~\ref{fig:activity_vs_class}. We note that our measured values have an upper limit of about 80\%, but that is not necessarily due to an absence of stars with $\sim$100\% of their time in decline. Rather, this points to the difficulty of obtaining spectra of those stars that are seldom (or never) at maximum light, so they are missing from the classified sample. A total of 47 stars are affected by missing spectral classifications, and those stars are included in Figure~\ref{fig:stats_histograms} but not in Figure~\ref{fig:activity_vs_class}. 

\begin{figure}
    \centering
    \includegraphics[width=\columnwidth]{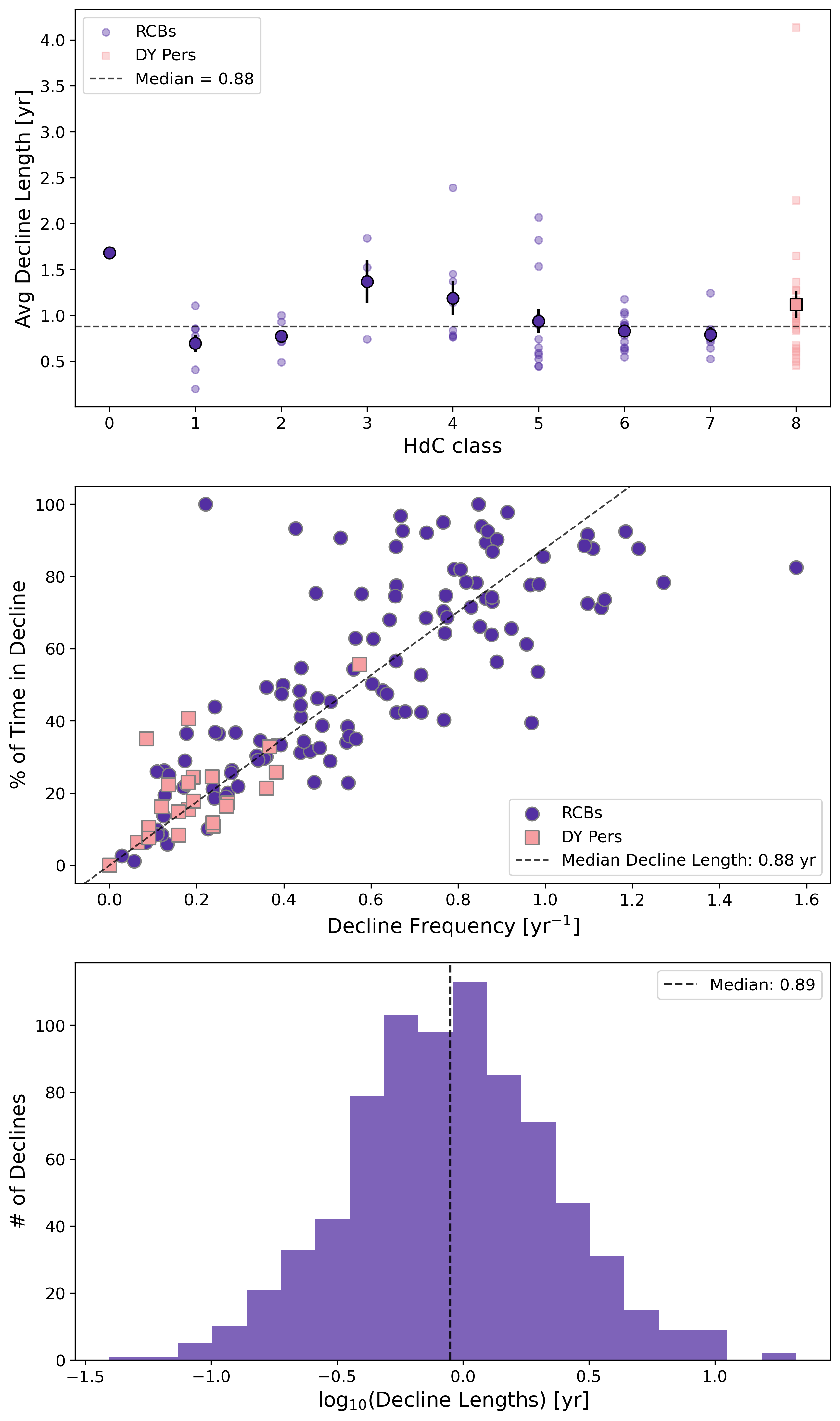}
    \caption{The top panel shows average decline lengths (calculated by dividing the percent of time in decline by the decline frequency) for each of the RCB stars (purple circles; Classes 0--7) and DY Per stars (peach squares; Class 8). The average decline length is plotted versus the HdC class, with a horizontal dashed grey line denoting the median of all points (0.88 years). The outlined markers show the mean of each HdC class, and the error bars mark the standard error on the mean. The central panel shows the percent of time in decline versus the decline frequency for each star, and the dashed grey line denotes the median of the average decline lengths as seen in the upper panel. The lower panel shows a histogram of all individual decline lengths in logarithmic space. The grey dashed line in this panel is the median of the distribution of individual decline lengths, i.e. not the same measurement as the upper two panels.}
    \label{fig:activity_stats_relations}
\end{figure}

These two activity indicators are related by the star's average decline length, and their distributions are so similar because this average decline length is similar from star to star. In fact, one can estimate the average decline length for each star by dividing the percent of time in decline by the decline frequency. In Figure~\ref{fig:activity_stats_relations}, we show the average decline length (calculated as described above) for each star as a function of spectral type (top panel) and the correlation between the two activity indicators (middle panel). In each of these representations, we indicate the mean of the average decline lengths, which is 0.88 years.

While the measurement of the activity indicators allows us to easily calculate the average decline length, we have also measured the duration of declines in each star directly. In Figure~\ref{fig:activity_stats_relations} (lower panel), we show the distribution of all individual decline lengths throughout the whole sample. Note that for this analysis we removed all nested declines, i.e., this only includes the duration of declines that begin while the star is at maximum. This will bias our distribution towards longer decline durations. Here we see that the total sample of decline lengths forms a log-normal distribution with a mean of 1.3 years and a median of 0.87 years. This distribution is consistent with the mean decline length calculated as the ratio of activity indicators (upper panel) previously. Therefore, we can conclude that a typical RCB decline is roughly 1 year long on average. However, declines longer than one year are not uncommon. As mentioned previously and shown in Figure~\ref{fig:rcrb-light curve}, R CrB recently had a 10 year long decline. In Figure~\ref{fig:activity_stats_relations} (lower panel), one can see that there are a significant number of declines longer than 5 years (0.7 on the x-axis). To be exact, out of the 1039 individual declines measured in this study, 30 declines or 3\% of total declines are longer than 5 years.

\begin{figure*}
    \centering
    \includegraphics[width=\textwidth]{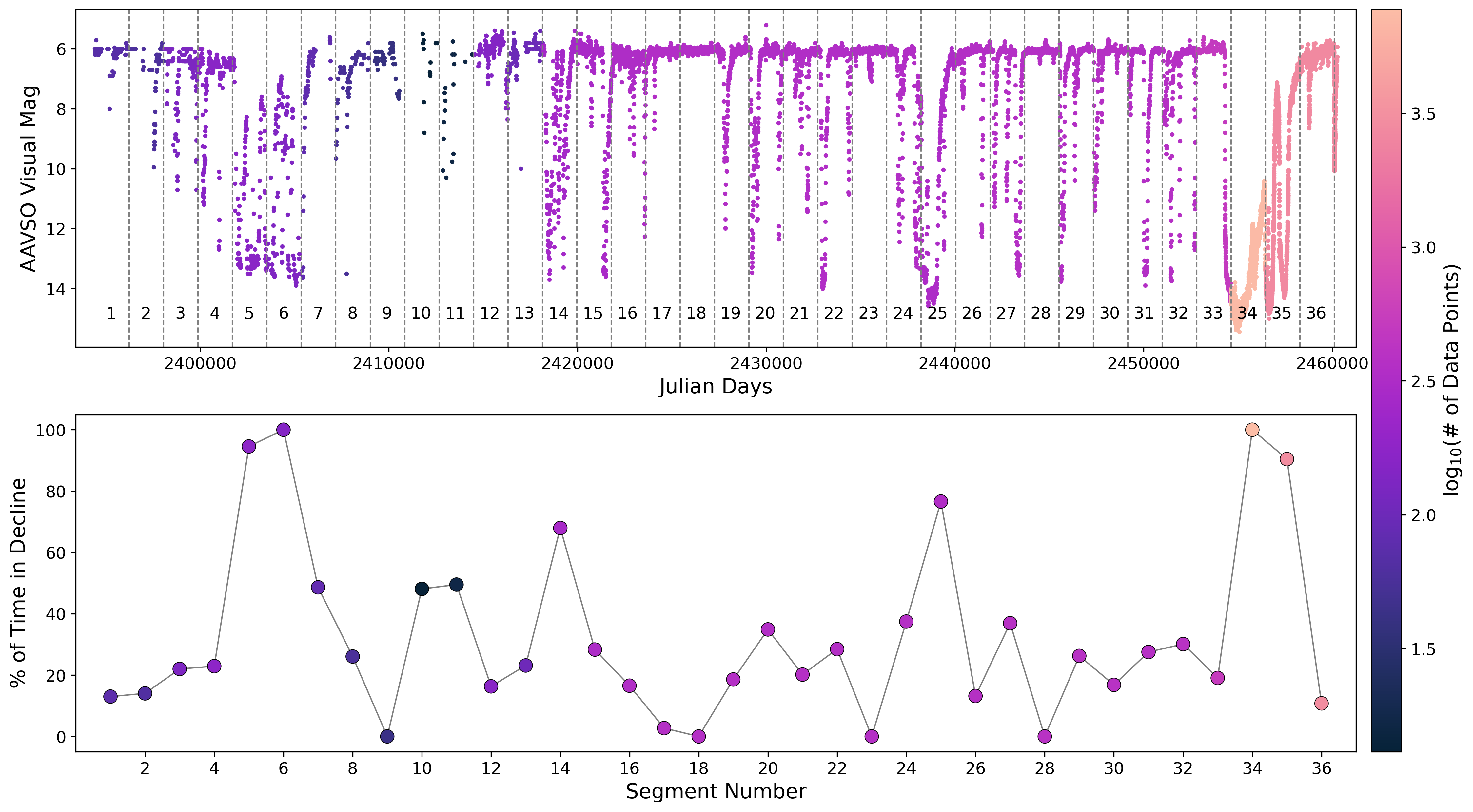}
    \caption{The upper panel shows the R CrB AAVSO light curve separated into 5 year segments and lower panel shows the percent of time each of those segments is spent in decline. The colored points show the logarithm of the number of data points in each of the 5 year segments as denoted by the colorbar, and serve as an indication of the precision of the measurement of the decline activity.}
   \label{fig:rcrb_segments}
\end{figure*}

While we report each individual star in our study as having one particular decline frequency and percentage of time spent in decline, it appears to be typical for RCB stars to have both active and inactive periods over long timescales, and thus these activity metrics will vary over time. Using the intensity of dust emission from K- and L-bands, \citet{1997MNRAS.285..317F} found that the dust production in a sample of 12 RCB stars rose and fell on a characteristic timescale of about 3 years. In Figure~\ref{fig:rcrb_segments}, we use the 125 year light curve of R CrB from AAVSO as an illustrative example. We cut R CrB's light curve into 5-year segments and calculated the percentage of time spent in decline for each of the segments, which we present as a function of time in the lower panel. We colour each 5-year segment to indicate the total number of data points it contains. The segments with the most data points will have the most reliably measured activity level. From this plot, we can see that R CrB has had many phases of high activity as well as a few phases of inactivity. If we had treated these segments independently, we would have calculated that R CrB spends an average of 20--30\% of its time in decline, which is consistent with the value of 32\% we obtained using the entire light curve. 

\subsection{Relation to H abundance}

\begin{figure}
    \centering
    \includegraphics[width=3.5in]{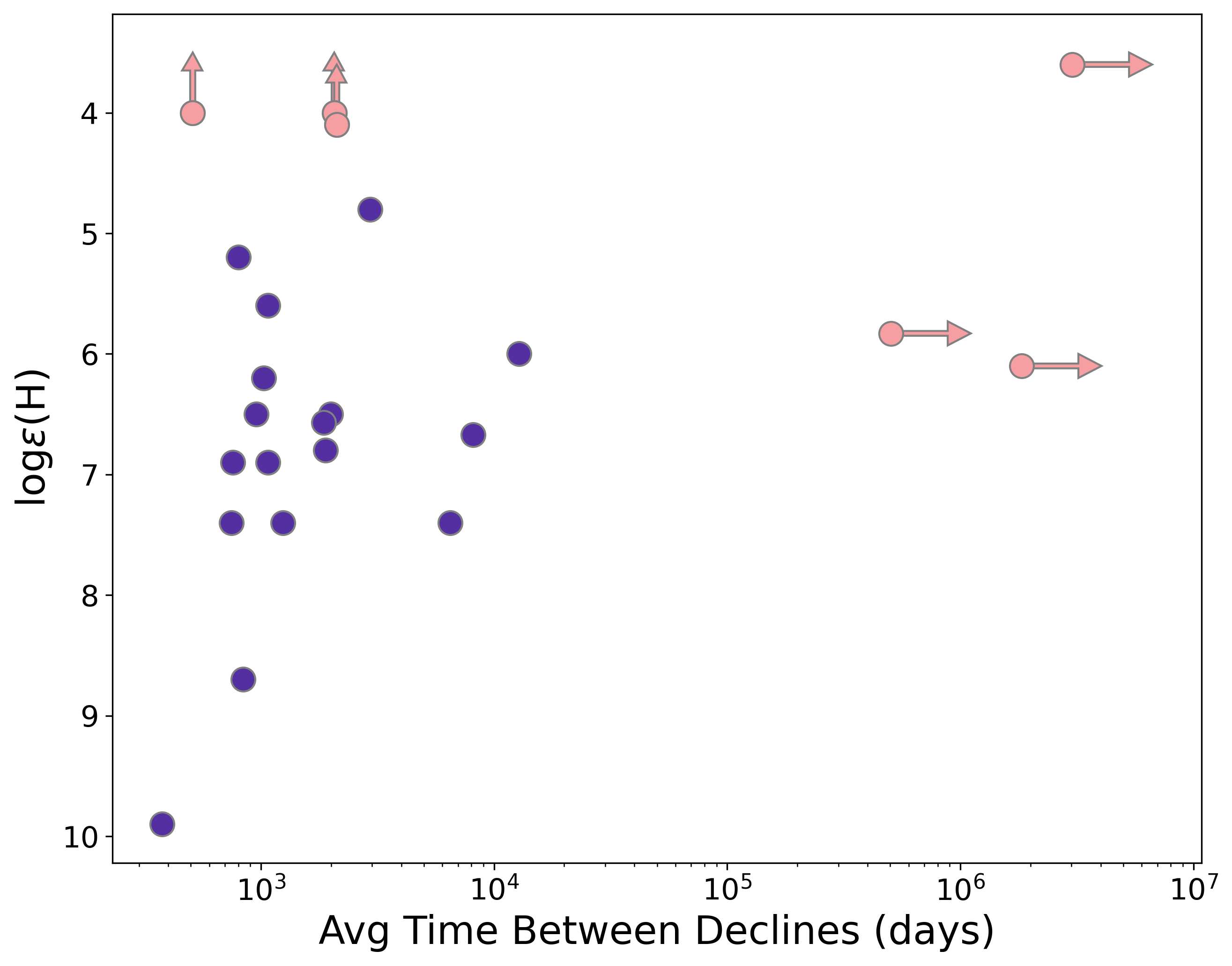}
\caption{Measured hydrogen abundance versus the average time between declines in log(days) for all RCB stars where these data are available. Note that the vertical axis is reversed, so stars with strong H abundances are lower in the plot. Peach points with arrows represent lower limits in either hydrogen abundance (for which the H abundance is too small to be accurately measured) or lower limits in the time between declines (for which the measured light curve does not exhibit declines, and a conservative lower limit is calculated as 1/3 of the total observation time of the light curve).}
\label{fig:habund}
\end{figure}

Figure 1 of \citet{1996AcA....46..325J} shows an apparent relationship between the average time between declines and hydrogen abundances of the RCB stars\footnote{Historically, HdC studies use a logarithmic abundance scale for an element X, where log$\epsilon$(X)~=~log$\frac{M_X}{\mu_X}$ + 12.15. $M_X$ is the mass fraction of element X and $\mu_X$ is the mean atomic mass of element X. log$\epsilon$(X) refers to the logarithm of the number of element `X' ions in a gas with 10$^{12.15}$ total ions.}. The average time between declines (log($\Delta$T$_{\rm fades}$) in their work) is equivalent to the reciprocal of the decline frequency reported in this work. In Figure~\ref{fig:habund}, we show an updated version with our decline frequencies and new abundances from the literature \citep{1979ApJ...233..205B,2000A&A...353..287A,2003PASP..115.1304R,2005BaltA..14..215K,2008ASPC..391...43K,2014PASP..126..813K,2017PASP..129j4202H}. We use upwards arrows to indicate stars with lower limits on H abundance. For stars with no measured declines, we use rightwards arrows at a location of 1/3 of the total observed time to denote a conservative lower limit for the time between declines. With the addition of more RCB stars, updated abundances, and longer baseline photometry, the apparent relationship found by \citet{1996AcA....46..325J} seems less likely. While the two abnormally H-rich stars are more actively producing dust, the bulk of the stars seem to have no apparent correlation between their H abundances and their dust production.

\subsection{DY Per Variables}

The known DY Per variables have similarities to cool RCB stars \citep{2009A&A...501..985T} and we have included them as class 8 in Figures~\ref{fig:activity_vs_class}~and~\ref{fig:activity_stats_relations}. It is clear that, although these two types of stars may appear similar spectroscopically, their decline activity levels are quite different. The DY Pers, even though they are cooler than the coolest RCBs, are significantly less active than the trend predicts. They only show an average of $\sim$0.2 declines per year and spend an average of 20\% of each year observed in decline. On the other hand, the average length of the DY Per declines is consistent with the RCB stars (see Figure~\ref{fig:activity_stats_relations}). Overall, this suggests a difference in the dust production mechanisms of the DY Pers and RCB stars.

It is important to keep in mind that only three of the known DY Per stars are Galactic, including DY Persei itself and two others (ASAS-DYPer-1 and ASAS-DYPer-2, \citealt{2013A&A...551A..77T}). All other known DY Per stars are members of the Magellanic Clouds \citep{2009A&A...501..985T}. This is presumably due to the lack of focused searches for DY Per stars in the Galaxy.
There is some evidence that the DY Per variables are created by a low-mass, solar-metallicity population of double white-dwarf binaries which merged recently \citep{Ruiter2019,Tisserand2024_3Ddistribution,Crawford2024_dlhdcmodels}. Spectral analysis of DY Per itself is difficult, and it is unclear whether its metallicity is solar or significantly lower \citep{Zacs2007_dypersolar,Yakovina2009_dyperabunds}. In addition to all the minor spectroscopic differences to the Magellanic Clouds, there is additionally a strong bias in the type of light curve data available for the Magellanic Clouds. The light curves available for these stars are significantly shorter than those available for most Galactic RCB stars, which may bias both their discovery and their decline statistics. 
As always, more data in the Magellanic Clouds would be extremely helpful for understanding these rare stars.


\section{Decline Morphologies}
\label{sec:decline_morph}

The morphologies of individual declines carry information on the dust properties of RCB stars. However, this information is difficult to measure for many stars because of gaps in the data, irregular data sampling, plate limits, and saturation, all of which affect the observed shape of a decline. In the following subsections, we discuss three aspects of decline morphologies: the depths of declines, the downward slope of the decline onset, and the shape of the decline recoveries. This analysis is not possible for all of the RCB stars in our sample and, in each case, we must be selective about which data we use for each star. Generally, this analysis is focussed on the brightest stars with the longest and best-sampled light curves.

\subsection{Decline Depths}
\label{subsec:depths}

The depth of a decline at its minimum may provide information on the relative projected size of the dust cloud compared to the angular size of the RCB star. However, during deep minima (8--9 mag) we may also see a contribution to the flux from the circumstellar reflection nebulae \citep{2018AJ....156..148M} or the stellar chromosphere \citep{1992ApJ...384L..19C}. There is evidence for this in the evolution of RCB decline spectra, which have been studied extensively \citep{1992ApJ...397..652C,1992AJ....103.1652W}. The flux from the dust-scattered light in R CrB has been estimated using polarimetric observations to be $\sim$10$^{-4}$ of the star's maximum brightness \citep{1992AJ....103.1652W,1993ASPC...45..115W}. Therefore, there may be an upper limit to the depth of an RCB decline, at which the RCB light curve will ``flatten'' due to dominant contributions from circumstellar dust scattering. 

R CrB is once again a useful case study, since it is the brightest RCB star at maximum ($V=6$ mag) and has been extensively observed by the AAVSO. Its 2007 decline (Figure~\ref{fig:rcrb-light curve}) was not only the longest, but was also nearly 1 mag deeper than any of its previously observed declines (Figure~\ref{fig:rcrb_segments}). This decline had a depth of 8.85 magnitudes (a 99.97\% reduction in flux), although this extreme value may reflect the availability of high-quality photometric data in the modern age. This very deep and long decline may indicate that the obscuring dust cloud expanded to eclipse some of the scattered light from dust around the star. It also shows that we should expect RCB stars to show declines up to $\sim$9 mag, provided the photometric precision is good enough. 

\begin{figure}
    \centering
    \includegraphics[width=\columnwidth]{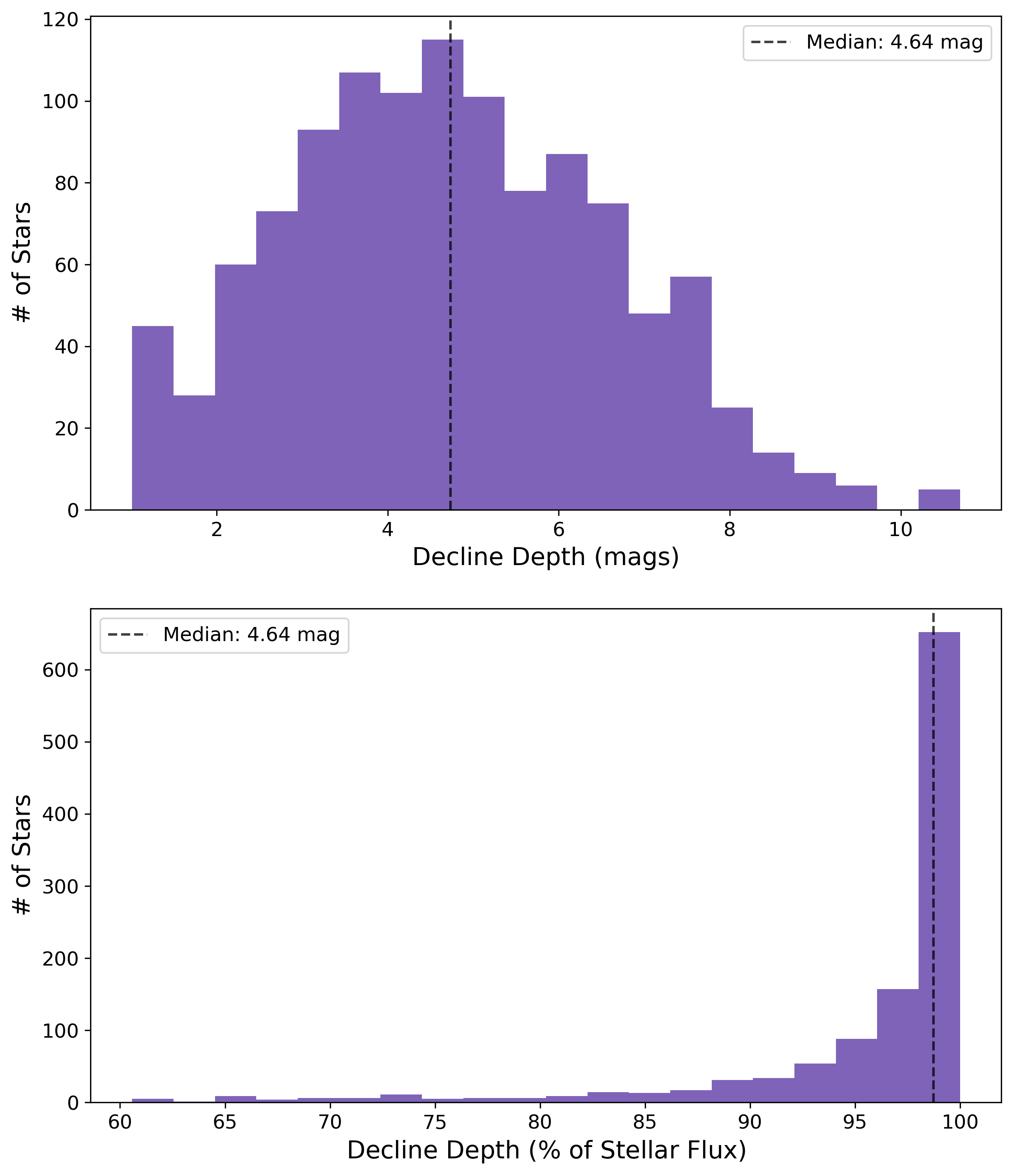}
    \caption{Histograms of the decline depths. The upper panel shows the decline depths in magnitudes, and the lower panel in percentage of total stellar flux at maximum light. The vertical dashed grey lines in both panels denote the median decline depth in magnitudes (4.64 mag).}
   \label{fig:depths_hist}
\end{figure}

As R CrB is so bright, a 9 magnitude decline is still easily observable at V=15 mag at minimum. However, many of the recently discovered RCB stars have maximum brightnesses of V$\sim$13, so that large telescopes are needed to measure their true minima. 
Keeping this is mind, we measured the depths for all observed declines, including those which are nested. We present the distributions in both magnitude and flux space in Figure~\ref{fig:depths_hist}. The median decline depth is 4.64 mag. This is representative of the average decline depth before the star hits the telescope faint limits. Notice as well that there are very few declines with overall depths $>$8 magnitudes. Extremely deep minima ($\sim$10 mag) are possibly indicative of poor photometric precision near the faint limit rather than a true detection of deep declines.

The lower panel of Figure~\ref{fig:depths_hist} shows decline depths in flux space, which paints a more realistic picture of the decline behaviour. Here, it is clear that the vast majority of RCB declines obscure more than 95\% of the total stellar flux. We point out that the upper limit of a 60\% reduction in flux comes from our requirement that a decline must be more than 1 magnitude in depth. Additionally, the median decline depth, 4.64 mag, corresponds to a reduction in stellar flux of $\sim$98\%. 

The decline depths are a particularly challenging feature of the dust to understand. In this analysis, we have not accounted for the fact that the decline depth also depends on the chosen passband, and the change in stellar photometric colour during a decline can tell us even more about the dust grains in the clouds. We leave such an analysis to our future work.

\subsection{Decline Slopes}
\label{subsec:slopes}

\begin{figure}
    \centering
    \includegraphics[width=\columnwidth]{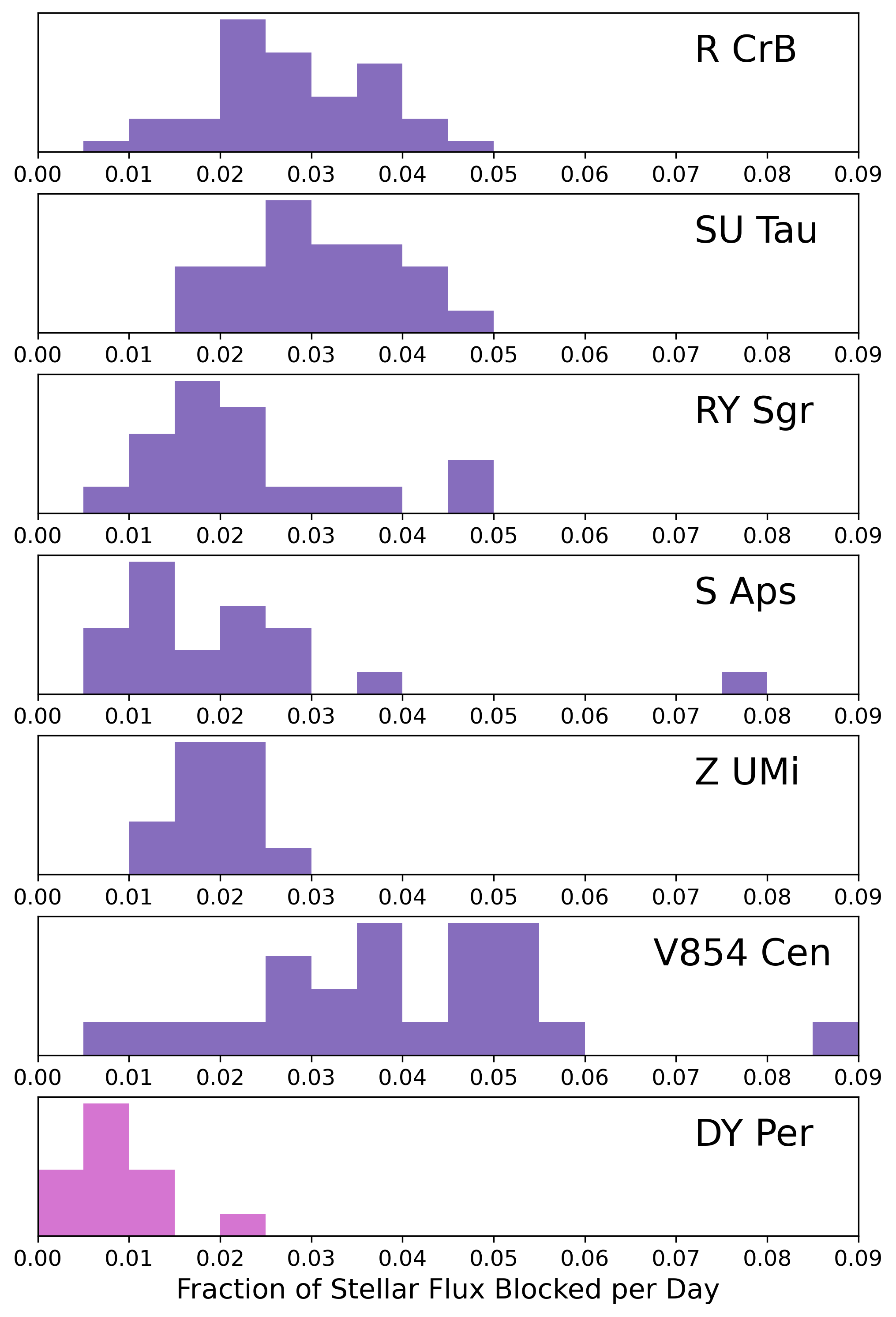}
    \caption{Histograms showing the decline slope of each individual decline for six RCB stars (purple: R CrB, SU Tau, RY Sgr, S Aps, Z UMi, V854 Cen) and DY Per (magenta). The slopes are measured in units of the fraction of stellar flux blocked per day (see Sec.~\ref{subsec:slopes}).}
   \label{fig:slopes}
\end{figure}

The shape of the decline onset probes the formation of the dust. Luckily, a simplistic measurement of the onset slope (assuming the decline onset is roughly linear) is relatively immune to measuring the true minimum of a decline. However, it does rely on the decline onset being well sampled in time. Additionally, single declines in single stars are not particularly informative, and we instead prefer to study the distribution of decline slopes for a few well-studied stars. Therefore, we chose to study 6 bright RCB stars with long baseline AAVSO visual light curves: R CrB, SU Tau, RY Sgr, S Aps, Z UMi, and V854 Cen, as well as DY Per. To measure decline onset slopes, we fitted a line from the decline onset time (which occurs at maximum brightness) to the decline minimum. In cases where the decline minimum was flat, we chose the minimum as the time closest to the onset. Additionally, we ignored nested declines and any declines where the onset or the minimum were not well sampled in time. We present the distributions in Figure~\ref{fig:slopes}, in units of the fraction of stellar flux blocked per day. 

The decline slopes measured here range from roughly 1\% to 6\% of the stellar flux blocked per day, with each RCB star having a slightly different distribution. In other words, if the average RCB decline blocks at least 95\% of the stellar flux, then it will take most stars between 15 and 95 days to reach minimum, with an average of 30--40 days. Some stars, such as Z UMi, tend to have slower declines, while SU Tau and V854 Cen have faster declines. The stars with slower declines seem to be cooler than those with faster declines, which is opposite to the naive expectation that dust should form faster around cooler stars.
Carbon chemistry models of pulsating RCB star atmospheres show that shock waves can produce non-LTE heating and cooling which lend themselves to dust production \citep{1986ASSL..128..151F,1996A&A...313..217W}. After the compression and re-expansion caused by the passage of the shock, the temperatures and densities required for nucleation of carbon dust can occur. These models predict typical timescales for this dust nucleation of 40--80 days, which is consistent with the formation timescales found here.

In the bottom panel of Figure~\ref{fig:slopes} we show the decline onset slopes for DY Persei. DY Per declines much more slowly than the RCB stars, with an average of less than 0.01\% of the flux blocked per day. This has been noticed previously, and it has also been said that DY Per's declines are ``more symmetrical'' than RCB star declines \citep{2001ApJ...554..298A}. The reason for this is not understood, and we do not observe enough declines in other DY Per type variables to know whether DY Persei is typical of the class.

\subsection{Decline Recoveries}
\label{subsec:recovery}

The shape of the decline recovery phase contains information about the distance of the dust from the star and the speed at which it moves away. In the simplest scenario, where circumstellar dust forms when gas reaches a distance consistent with the condensation temperature ($\sim$1100 K for amorphous carbon; \citealt{1984A&A...132..163G}), a 7000-K RCB star with $M_V=-5$ should form dust at $\sim50 R_\star$. Similarly, a 5000-K RCB star with $M_V=-3$ should form dust at $\sim20 R_\star$. However, these estimates assume a stellar wind, for which there is no evidence \citep[][and references therein]{1996PASP..108..225C}. Instead, the dust from RCB stars is accelerated by radiation pressure, and this dust can reach speeds of $\sim$400 km s$^{-1}$ \citep{1993ASPC...45..115W,2013AJ....146...23C} 

We can model the recovery phase of an isolated decline using a simple model. If it is assumed that once a dust cloud forms near the star, each dust grain moves radially away from the star, then amount of extinction varies as $d^{-2}$ where $d$ is the distance from the star. If the optical extinction is measured at two times, $t_a$ and $t_b$, at distances $d_a$ and $d_b$, then
\begin{equation}
\label{eqn:dust_radiation}
\  d_a= \frac{vt}{(\frac{A_{va}}{A_{vb}})^{1/2}-1}. 
\end{equation}
Here, $t=t_b-t_a$, and $A_{va}$ and $A_{vb}$ are the extinctions (in other words, the depths in magnitudes below maximum light) at $d_a$ and $d_b$ \citep{1993ASPC...45..115W}. 

\begin{figure*}
    \centering
    \includegraphics[width=\textwidth]{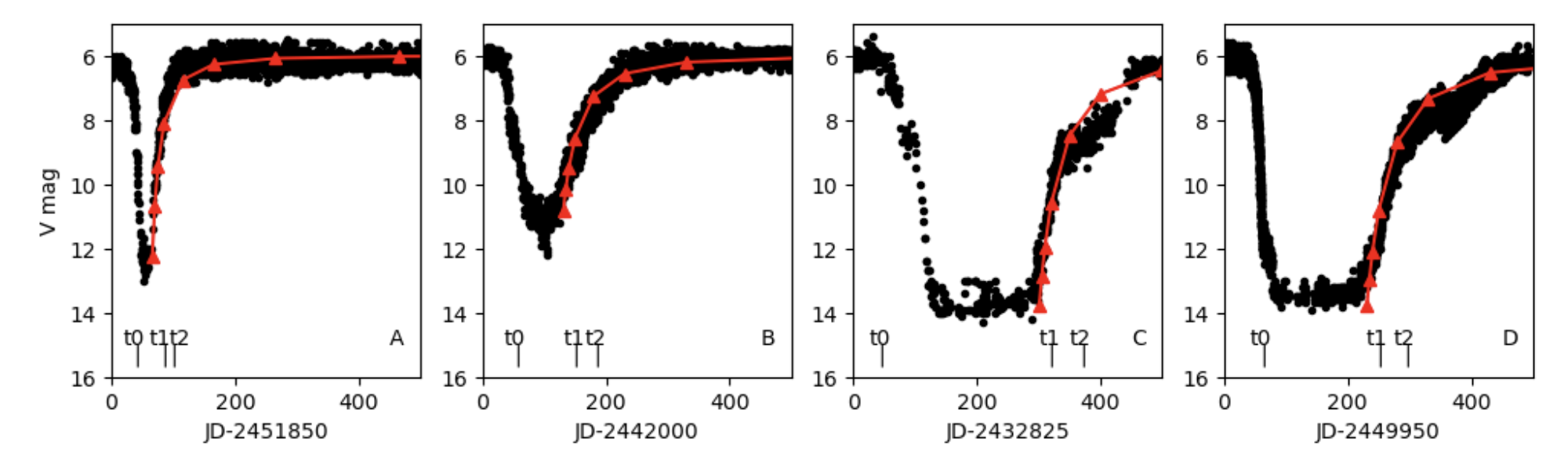}
    \caption{Four declines of R CrB (black points) with models (red lines and triangle markers) for the recovery to maximum light after the decline. The models assume that the dust is moving radially away from the star at 400 km s$^{-1}$ following Equation~\ref{eqn:dust_radiation}.
    t$_0$ denotes the time of the beginning of the decline, t$_1$ denotes the time of the beginning of the recovery, and t$_2$ denotes the time when the star reaches A$_V$ = 2 mag on the way up. 
    The parameters for each model are listed in Table~\ref{tab:decline_recovery_model}.}
    \label{fig:model1}
\end{figure*}

In Figure~\ref{fig:model1}, we show fits to four decline recoveries of R CrB, using Equation~\ref{eqn:dust_radiation}. 
The fit parameters for each decline are listed in Table~\ref{tab:decline_recovery_model}. 
The times, measured and marked in Figure~\ref{fig:model1}, are $t_0$, the beginning of the decline, $t_1$, the time when recovery begins, and $t_2$, the time when the star has recovered to 2 mag below maximum. Below we discuss two aspects learned from these fits.

First, the dust around R CrB has been measured to have velocities of $\sim$400 km $s^{-1}$ \citep{1993ASPC...45..115W,2013AJ....146...23C}, which are much larger than the fitted average velocity between the formation ($t_0$) and the beginning of recovery ($t_1$). We interpret this discrepancy as the impact of the time it takes to accelerate the dust to 400 km s$^{-1}$. However, models of dust acceleration by radiation pressure predict that this acceleration should be rapid \citep{1976A&A....52..245B,1976ApJ...203..552M,2013AJ....146...23C}. 

Second, we see that the amount of time between the decline onset ($t_0$) to the decline recovery ($t_1$) varies for each decline. Even though the decline onset speeds are very similar for each decline, the time spent at minimum is variable. This could indicate continuous dust production during the decline. In the model fit, it is represented as a change in the distance of the dust from the star when the recovery begins. This distance also affects the steepness of the recovery and thus how long the recovery phase takes to return to maximum light. In other words, the declines which spend less time at minimum also take less time to fully recover to maximum light, due to the dust being closer to the star when the recovery begins.

\begin{table*}
	\caption{Decline Recovery Parameters used in Fig~\ref{fig:model1}}
         \label{tab:decline_recovery_model}
	\def\arraystretch{1.15}
	\centering
	\begin{tabular}{ccccc}
		\hline
  Model	&	t$_1$-t$_0$	&	t$_2$-t$_1$	&	V$_{t_1-t_0}$	&	d$_{t_1}$	\\
  &(d)&(d)&(km s$^{-1}$)&(R$_*$)\\
  \hline
  A&45 &14 &207& 20\\
  B& 95 &35 &196& 40\\
  C&275& 50 &85& 50\\
D&188 &45 &136& 55\\
		\hline
	\end{tabular}
\end{table*}

\section{Decline Periodicity}
\label{sec:decline_onsets}

\citet{1977IBVS.1277....1P}, \citet{1999AJ....117.3007L}, \citet{2007MNRAS.375..301C}, and \citet{2018JAVSO..46..127P} have suggested that the onset times of declines in RCB stars are intrinsically tied to their pulsation phase. This does not necessarily imply that an RCB star will produce dust in each pulsation phase, instead that dust declines occur at certain phases of the pulsation cycle. In practice, each of the aforementioned works have provided linear ephemerides which predict the decline onset times for a small number of RCB stars. These ephemerides have the form ${\rm JD}_n = {\rm JD}_0 + n T_{\rm puls}$, where JD$_n$ is the onset time of a decline that occurs in the $n$-th pulsation cycle, $T_{\rm puls}$ is the pulsation period and JD$_0$ is a zero-point. They predict that declines should occur for integer multiples of $n$. However, some of these works reported differences of observed declines from the ephemerides of up to 10 days, which is roughly 25\% of the average pulsation period of 40 days \citep{2007MNRAS.375..301C}. 

There is some evidence that dust formation may be tied to the pulsation of the RCB stars. The best example are the pulsations of RY Sgr, which have the largest amplitude among known RCB star and which trigger shock waves in the stellar envelope \citep{Danziger1963_rysgrshocks,Cottrell1982_rysgrshocks,Lawson1991_rysgrshocks,Clayton1994_rysgrshocks}. This fits with the \citet{1996A&A...313..217W} model of dust formation in RCB stars. On the other hand, coherent pulsation periods with amplitudes comparable to RY Sgr have not been observed in any other RCB stars. For example, R CrB itself, which is clearly variable at maximum light, has appeared to have a varying pulsation period \citep{2004JAVSO..33...27P}, and long-term RV measurements have shown no evidence of coherent pulsations \citep{Feast2019_rcrbRVs}. 
These studies indicate that the variations of R CrB are probably stochastic in nature \citep{1997MNRAS.285..339F,Feast2019_rcrbRVs}. Rather than being caused by pulsations, they may be due to surface convective cells, as has also been suggested for red supergiants \citep{1975ApJ...195..137S,2006MNRAS.372.1721K,2022ApJ...929..156G}.
By extension, the pulsation periods for many RCB stars are not well constrained, and have larger uncertainties than have been quoted by the ephemerides. This uncertainty exposes a significant weakness in the conclusions drawn by the aforementioned works which assume a specific pulsational period for individual RCB stars, which was previously discussed in \citet{Feast2019_rcrbRVs}.

In this work, we have measured the decline onset times for the five stars studied by \citet{2007MNRAS.375..301C}. However, using all available AAVSO data, we did not find indications of decline onset periodicity when inspecting the Fourier spectrum of the decline onsets. Even when restricting to only the same period of time measured by \citet{2007MNRAS.375..301C}, we do not see significant signals in the amplitude spectra. When we compare our determined onset times to those of Crause et al., we find that the times vary with an average scatter of roughly 5 days, which can most likely be attributed to the differences in manual decline detection. 

\begin{figure}
    \centering
    \includegraphics[width=\columnwidth]{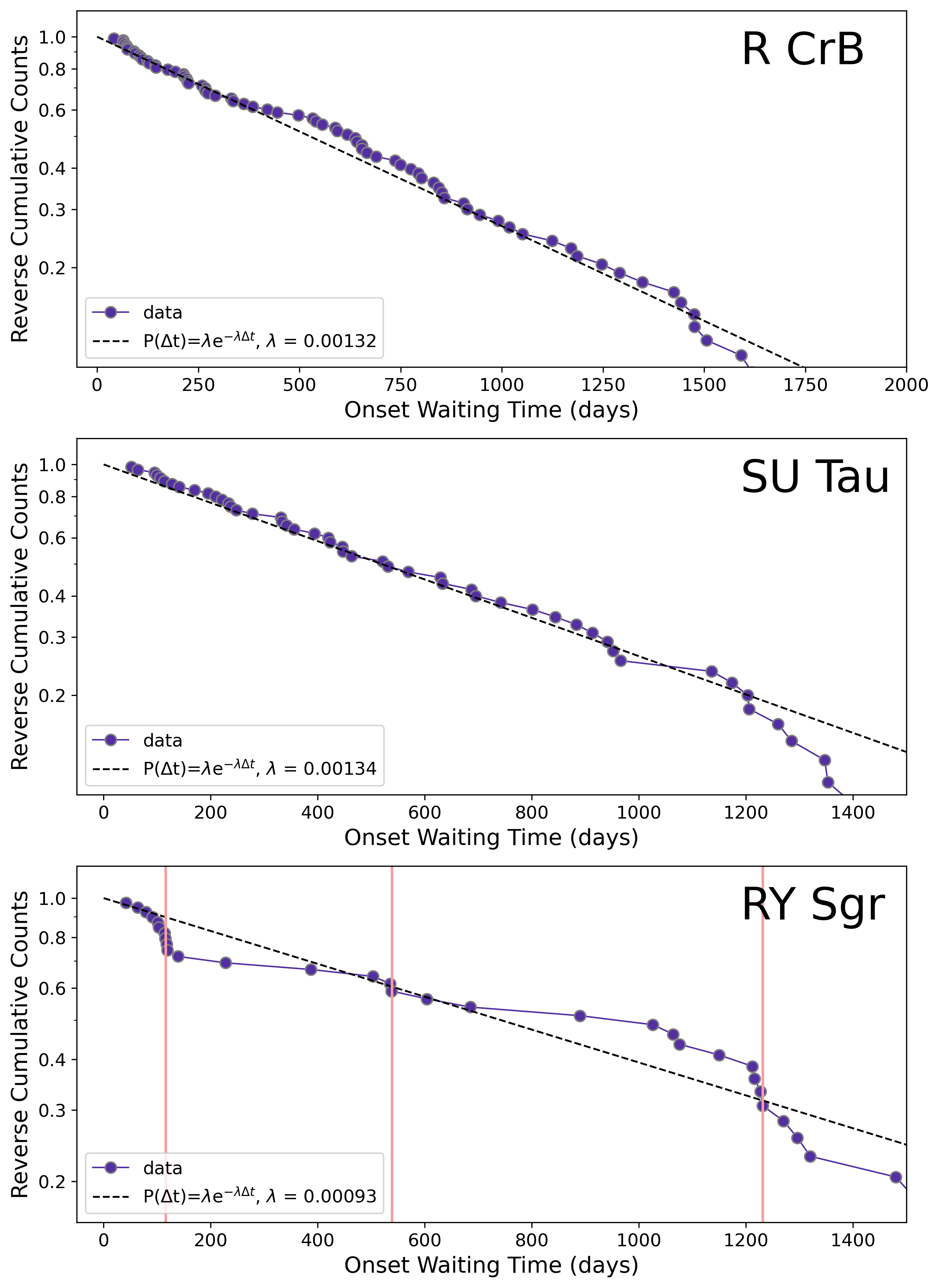}
    \caption{The waiting time distribution (see e.g. \citealt{Wheatland2000_solarflareoccurrence}) for 3 RCB stars (R CrB, SU Tau, and RY Sgr) in purple markers compared to a Poisson distribution (dashed grey line) where the Poisson rates are adopted as the decline frequency calculated in Section~\ref{sec:decline_rate}, rather than fitted to the data. For RY Sgr in the lower panel, we mark three regions of the distribution with peach vertical lines where there are many decline onsets with similar wait times. These three lines are at times $3T$, $14T$, and $32T$ where $T$ is RY Sgr's pulsation period (38.46 days).}
    \label{fig:wtds}
\end{figure}

Rather than attempting to define our own decline onset ephemeris, we instead explored the idea that the declines may occur as a Poisson process. This analysis is directly analogous to that of solar flare occurrence rates (see e.g. \citealt{Wheatland2000_solarflareoccurrence}). We chose 3 example stars which have both long-baseline and well-sampled AAVSO light curves and then calculate their waiting time distributions (WTDs). A `waiting time' is defined as the time interval between subsequent decline events, and calculated as the pairwise difference in onset times, which is then presented as a cumulative distribution. In Figure~\ref{fig:wtds} we show the WTDs for SU Tau, R CrB, and RY Sgr. The observed data are shown in purple markers. In the dashed grey lines, we show a Poisson distribution $P(\Delta t) = \lambda e^{-\lambda\Delta t}$, which assumes a constant rate $\lambda$. We assume $\lambda$ is the frequency of declines (see Section~\ref{sec:decline_rate}), rather than performing a fit to the data. Here we see remarkable agreement with the Poisson distribution for SU Tau and R CrB, which implies their dust production is driven by a similarly stochastic process such as surface convection.

However, RY Sgr shows weaker agreement than the other two stars. In fact, RY Sgr seems to have some periodicity, i.e., there are multiple declines that occur at very similar delay times. On the RY Sgr panel of Figure~\ref{fig:wtds}, we have included three vertical lines to indicate where there are delay time build-ups. These three lines are at $3T$, $14T$, and $32T$ where $T$ is 38.46 days, the measured pulsation period for RY Sgr \citep{1994ApJ...432..785C,1997MNRAS.285..266L}. The build-up of decline onsets at integer multiples of the pulsation period aligns with the predictions from  other studies of RY Sgr \citep{1977IBVS.1277....1P,2007MNRAS.375..301C}. This seems to imply that a major component of the dust production for RY Sgr is driven by its especially high-amplitude pulsations \citep{1997MNRAS.285..266L}. RY Sgr is the best candidate for pulsation-driven dust formation. However, there may be a stochastic (Poisson) component in the dust production of RY Sgr as well.

The actual mechanism for RCB dust production remains uncertain. From this analysis, we see conflicting evidence on whether or not there is a relationship between the dust formation and the pulsation cycle of RCB stars. It seems additionally that the dust production may vary between different stars, dependent on their oscillation or convective properties. Even in RY Sgr, which shows signatures of a decline-pulsation relationship, there may be a Poisson-distributed component to the dust production as well.

\section{Conclusions}
\label{sec:conclusions}

This study of RCB declines is the largest to date. With data from 162 RCB stars and multiple photometric sources, we have examined RCB decline activity over the longest possible timescales. From this extensive dataset, we have learned the following:

1. We have confirmed the prediction from \citet{2024A&A...684A.130T} that cool RCB stars decline more often and spend more time in decline overall than warm RCB stars. However, there are well-known exceptions to this, such as XX Cam, a cool star with only one observed decline in $\sim$100 years, and V854 Cen, a warm 7th magnitude star that went undiscovered until the 1980's due to a long-lasting deep decline \citep{1986IAUC.4233....3M}.

2. We find that the average decline length for a single RCB decline is roughly one year. This is consistent regardless of if the average decline length is measured using the average decline rates per star or by examining individual declines. Examining individual declines also reveals that the decline lengths follow a log-normal distribution. Why the declines follow such a distribution or why they are distributed around one year is unclear, though it may indicate an unaccounted for observational bias.

3. Within any single star, there are large variations in the decline activity over $\sim$few year timescales. For example, R CrB has multiple 5-year segments of time where it has no declines and multiple segments where it spends more than 50\% of observed time in decline. This is consistent with results from \citet{1997MNRAS.285..317F} which state that dust production in RCBs varies on a timescale of about 3 years.

4. With more decline and abundance data, the relationship between H and decline activity suggested by \citet{1996AcA....46..325J} has weakened. Barring the exceptionally large H abundances of V854 Cen and V CrA, the RCB stars span the entire range of decline frequencies and of H abundances with little correlation.

5. We do not measure the true minima of most RCB declines. However, using the available data (which is biased toward shallower declines), we find that most declines obscure over 95\% of the stellar flux. The only star for which we are confident in measuring the true decline minimum is R CrB, and its largest decline is 8.85 magnitudes deep (a 99.97\% reduction in flux, or 0.0003 of the maximum brightness). This deep decline flattens at the minimum, suggesting that the flux may be dominated by scattered light from surrounding dust, which is estimated to be $\sim$10$^{-4}$ of the star's peak brightness \citep{1992AJ....103.1652W,1993ASPC...45..115W}.

6. RCB stars show a (small) range of different decline onset slopes, which probe the characteristic dust nucleation time for each star. The average RCB decline takes 30-40 days to reach its minimum, which is consistent with the 40-80 day timescales predicted by carbon chemistry dust production models \citep{1986ASSL..128..151F,1996A&A...313..217W}. 

7. The recovery light curves can be fit using a radially moving dust with a dust velocity of 400 km/s, as measured by \citep{1993ASPC...45..115W,2013AJ....146...23C}. However, the dust radius required for the fits in combination with the time elapsed from decline onset shows that it takes longer than expected for the dust to accelerate to these velocities.

8. We show that R CrB and SU Tau's decline waiting time distributions follow a Poisson distribution where the Poisson rate is the average decline frequency. This indicates that the declines in these stars occur as a Poisson process. However, RY Sgr, which is the only RCB star with direct evidence of pulsation-driven shock waves in its atmosphere \citep{Danziger1963_rysgrshocks,Cottrell1982_rysgrshocks,Lawson1991_rysgrshocks,Clayton1994_rysgrshocks}, shows indications of its declines occurring at integer multiples of the pulsation period, although there are many declines of RY Sgr which do not seem related to the pulsation period. Considering there is no published evidence for long-term coherent pulsations in R CrB \citep{Feast2019_rcrbRVs}, the dust production in this star is likely driven by a stochastic process, such as convection. The WTDs seem to suggest there may be two different dust production mechanisms in the RCB stars.

9. Despite the spectroscopic similarities between RCB stars and DY Per stars, the DY Per stars seem to have quite different dust production properties. The DY Pers decline less frequently than would be predicted by their temperatures, though the length of their average decline is the same as RCBs. Additionally, DY Persei's declines take roughly twice as long to reach their minima as RCB star declines do. 

Dust formation around RCB stars remains a poorly understood phenomenon, and with this work we hope to revitalize interest in the topic. Current data suggests that RCB stars produce dust ``puffs" near the stellar surface ($\sim$2R$_\star$), which radiate outwards from the star at a speed of $\sim$400 km/s. If these puffs occur in the line of sight, they obscure the photosphere as they expand and eventually cover the circumstellar emission regions, before growing to a size such that they become optically thin and the photosphere re-emerges \citep[][and references therein]{1996PASP..108..225C}. These dust puffs seem to occur preferentially in certain regions of the star, likely the equator, such that some stars do not exhibit visible dust declines despite strong evidence of circumstellar dust. The dust itself consists of amorphous carbon according to UV extinction measurements, and its nucleation timescale is consistent with carbon dust chemistry models. 

At least three major questions remain in our understanding of RCB dust. The first is the dust production mechanism itself. We have shown that different RCB stars may have different dust production mechanisms. While the dust production in some stars follows a Poisson process, other stars seem to be driven by stellar pulsations. New 3D models of luminous stars, like those in \citet{Freytag2023_agbdustmodel3D}, could shed light on how convective cells and pulsations influence dust formation. Additionally, the study of similar erratic dust-producing stars such as some C-rich Miras may guide future work on the relationship between pulsations and dust production \citep[e.g.][]{1997MNRAS.288..512W,2003MNRAS.346..878F,2006MNRAS.369..751W}. The second question is why the RCB stars have a preferred location for their dust production. The third question, which we have not touched on significantly in this work, is the difference between the dust producing RCB stars and their dustless counterparts, the dLHdCs. There are only a few known spectroscopic differences between the two types of stars \citep{2022A&A...667A..84K,2022A&A...667A..83T}, which can be explained using different masses of progenitor stars \citep{Crawford2024_dlhdcmodels}, but we do not understand enough about the dust production to infer how these differences could prevent declines in dLHdCs.

Despite the long-baseline observations in this study, most RCB light curves are shorter than 4 years, providing an incomplete view of RCB dust activity. Long-baseline continuous observations are crucial for our work, and thus certain upcoming missions will be invaluable to us. The Legacy Survey of Space and Time (LSST) on the Vera Rubin Observatory \citep{LSST} plans to cover the entire Southern sky for 10 years, enabling us to monitor the majority of RCB stars. The Planetary Transits and Oscillations of Stars (PLATO) mission \citep{Plato1,Plato2} also shows promise, as it will feature continuous monitoring of large patches of sky in the Northern and Southern hemispheres, and will hopefully feature a few RCB stars in its footprint. In the interim, we will continue to monitor the RCB stars for new and interesting behaviour, and hope to continue unveiling new characteristics.

\section*{Acknowledgements}

CC and TB gratefully acknowledge support from the Australian Research Council through Discovery Project DP210103119 and Laureate Fellowship FL220100117. C.-U. Lee acknowledges support from the Korea Astronomy and Space Science Institute under the R\&D program (Project No. 2024-1-832-01) supervised by the Ministry of Science and ICT.

The authors thank Elise McCullison, Sara Sayer, and Aidan English, who worked on this project as undergraduates at Louisiana State University. We would like to thank James Crowley and Michael Wheatland for discussions on solar flare occurrence rates and wait time distributions. We acknowledge with thanks the variable star observations from the AAVSO International Database contributed by observers worldwide and used in this research. 

This research has made use of the KMTNet system operated by the Korea Astronomy and Space Science Institute (KASI) at three host sites of CTIO in Chile, SAAO in South Africa, and SSO in Australia.
Data transfer from the host site to KASI was supported by the Korea Research Environment Open NETwork (KREONET).

This work was partially based on observations obtained with the Samuel Oschin Telescope 48-inch and the 60-inch Telescope at the Palomar
Observatory as part of the Zwicky Transient Facility project. ZTF is supported by the National Science Foundation under Grants
No. AST-1440341 and AST-2034437 and a collaboration including current partners Caltech, IPAC, the Oskar Klein Center at
Stockholm University, the University of Maryland, University of California, Berkeley , the University of Wisconsin at Milwaukee,
University of Warwick, Ruhr University, Cornell University, Northwestern University and Drexel University. Operations are
conducted by COO, IPAC, and UW.

This work made use of several publicly available {\tt python} packages: {\tt astropy} \citep{astropy:2013,astropy:2018}, 
{\tt matplotlib} \citep{matplotlib2007}, 
{\tt numpy} \citep{numpy2020}, and 
{\tt scipy} \citep{scipy2020}.

\section*{Data Availability}

Our full light curve data for all RCB stars are available online at https://dreams.anu.edu.au/monitoring/. The AAVSO data is publicly available through www.aavso.org. A table of all the measured activity metrics is available on Zenodo. The data from each individual decline detection can be requested from the corresponding author.

\typeout{}
\bibliographystyle{mnras}
\bibliography{refs}{}

\begin{thebibliography}{}
\makeatletter
\relax
\def\mn@urlcharsother{\let\do\@makeother \do\$\do\&\do\#\do\^\do\_\do\%\do\~}
\def\mn@doi{\begingroup\mn@urlcharsother \@ifnextchar [ {\mn@doi@} {\mn@doi@[]}}
\def\mn@doi@[#1]#2{\def\@tempa{#1}\ifx\@tempa\@empty \href {http://dx.doi.org/#2} {doi:#2}\else \href {http://dx.doi.org/#2} {#1}\fi \endgroup}
\def\mn@eprint#1#2{\mn@eprint@#1:#2::\@nil}
\def\mn@eprint@arXiv#1{\href {http://arxiv.org/abs/#1} {{\tt arXiv:#1}}}
\def\mn@eprint@dblp#1{\href {http://dblp.uni-trier.de/rec/bibtex/#1.xml} {dblp:#1}}
\def\mn@eprint@#1:#2:#3:#4\@nil{\def\@tempa {#1}\def\@tempb {#2}\def\@tempc {#3}\ifx \@tempc \@empty \let \@tempc \@tempb \let \@tempb \@tempa \fi \ifx \@tempb \@empty \def\@tempb {arXiv}\fi \@ifundefined {mn@eprint@\@tempb}{\@tempb:\@tempc}{\expandafter \expandafter \csname mn@eprint@\@tempb\endcsname \expandafter{\@tempc}}}

\bibitem[\protect\citeauthoryear{{Alcock} et~al.,}{{Alcock} et~al.}{2001}]{2001ApJ...554..298A}
{Alcock} C.,  et~al., 2001, \mn@doi [\apj] {10.1086/321369}, \href {https://ui.adsabs.harvard.edu/abs/2001ApJ...554..298A} {554, 298}

\bibitem[\protect\citeauthoryear{{Asplund}, {Gustafsson}, {Lambert}  \& {Rao}}{{Asplund} et~al.}{2000}]{2000A&A...353..287A}
{Asplund} M.,  {Gustafsson} B.,  {Lambert} D.~L.,   {Rao} N.~K.,  2000, \aap, \href {https://ui.adsabs.harvard.edu/abs/2000A&A...353..287A} {353, 287}

\bibitem[\protect\citeauthoryear{{Astropy Collaboration}}{{Astropy Collaboration}}{2013}]{astropy:2013}
{Astropy Collaboration} 2013, \mn@doi [\aap] {10.1051/0004-6361/201322068}, \href {http://adsabs.harvard.edu/abs/2013A%26A...558A..33A} {558, A33}

\bibitem[\protect\citeauthoryear{{Astropy Collaboration}}{{Astropy Collaboration}}{2018}]{astropy:2018}
{Astropy Collaboration} 2018, \mn@doi [\aj] {10.3847/1538-3881/aabc4f}, \href {https://ui.adsabs.harvard.edu/abs/2018AJ....156..123A} {156, 123}

\bibitem[\protect\citeauthoryear{{Bauer} \& {de Kat}}{{Bauer} \& {de Kat}}{1997}]{1997NIMPA.387..286B}
{Bauer} F.,  {de Kat} J.,  1997, \mn@doi [Nuclear Instruments and Methods in Physics Research A] {10.1016/S0168-9002(96)01009-1}, \href {https://ui.adsabs.harvard.edu/abs/1997NIMPA.387..286B} {387, 286}

\bibitem[\protect\citeauthoryear{{Bergeat}, {Lefevre}, {Kandel}, {Lunel}  \& {Sibille}}{{Bergeat} et~al.}{1976}]{1976A&A....52..245B}
{Bergeat} J.,  {Lefevre} J.,  {Kandel} R.,  {Lunel} M.,   {Sibille} F.,  1976, \aap, \href {https://ui.adsabs.harvard.edu/abs/1976A&A....52..245B} {52, 245}

\bibitem[\protect\citeauthoryear{{Bhowmick}, {Pandey}, {Joshi}  \& {Ashok}}{{Bhowmick} et~al.}{2018}]{2018ApJ...854..140B}
{Bhowmick} A.,  {Pandey} G.,  {Joshi} V.,   {Ashok} N.~M.,  2018, \mn@doi [\apj] {10.3847/1538-4357/aaaae4}, \href {https://ui.adsabs.harvard.edu/abs/2018ApJ...854..140B} {854, 140}

\bibitem[\protect\citeauthoryear{{Bond}, {Luck}  \& {Newman}}{{Bond} et~al.}{1979}]{1979ApJ...233..205B}
{Bond} H.~E.,  {Luck} R.~E.,   {Newman} M.~J.,  1979, \mn@doi [\apj] {10.1086/157382}, \href {https://ui.adsabs.harvard.edu/abs/1979ApJ...233..205B} {233, 205}

\bibitem[\protect\citeauthoryear{{Clayton}}{{Clayton}}{1996}]{1996PASP..108..225C}
{Clayton} G.~C.,  1996, \mn@doi [\pasp] {10.1086/133715}, \href {https://ui.adsabs.harvard.edu/abs/1996PASP..108..225C} {108, 225}

\bibitem[\protect\citeauthoryear{{Clayton}}{{Clayton}}{2012}]{2012JAVSO..40..539C}
{Clayton} G.~C.,  2012, \mn@doi [\jaavso] {10.48550/arXiv.1206.3448}, \href {https://ui.adsabs.harvard.edu/abs/2012JAVSO..40..539C} {40, 539}

\bibitem[\protect\citeauthoryear{{Clayton}, {Whitney}, {Stanford}, {Drilling}  \& {Judge}}{{Clayton} et~al.}{1992a}]{1992ApJ...384L..19C}
{Clayton} G.~C.,  {Whitney} B.~A.,  {Stanford} S.~A.,  {Drilling} J.~S.,   {Judge} P.~G.,  1992a, \mn@doi [\apjl] {10.1086/186253}, \href {https://ui.adsabs.harvard.edu/abs/1992ApJ...384L..19C} {384, L19}

\bibitem[\protect\citeauthoryear{{Clayton}, {Whitney}, {Stanford}  \& {Drilling}}{{Clayton} et~al.}{1992b}]{1992ApJ...397..652C}
{Clayton} G.~C.,  {Whitney} B.~A.,  {Stanford} S.~A.,   {Drilling} J.~S.,  1992b, \mn@doi [\apj] {10.1086/171821}, \href {https://ui.adsabs.harvard.edu/abs/1992ApJ...397..652C} {397, 652}

\bibitem[\protect\citeauthoryear{{Clayton}, {Whitney}  \& {Mattei}}{{Clayton} et~al.}{1993}]{1993PASP..105..832C}
{Clayton} G.~C.,  {Whitney} B.~A.,   {Mattei} J.~A.,  1993, \mn@doi [\pasp] {10.1086/133240}, \href {https://ui.adsabs.harvard.edu/abs/1993PASP..105..832C} {105, 832}

\bibitem[\protect\citeauthoryear{{Clayton}, {Lawson}, {Cottrell}, {Whitney}, {Stanford}  \& {de Ruyter}}{{Clayton} et~al.}{1994a}]{Clayton1994_rysgrshocks}
{Clayton} G.~C.,  {Lawson} W.~A.,  {Cottrell} P.~L.,  {Whitney} B.~A.,  {Stanford} S.~A.,   {de Ruyter} F.,  1994a, \mn@doi [\apj] {10.1086/174616}, \href {https://ui.adsabs.harvard.edu/abs/1994ApJ...432..785C} {432, 785}

\bibitem[\protect\citeauthoryear{{Clayton}, {Lawson}, {Cottrell}, {Whitney}, {Stanford}  \& {de Ruyter}}{{Clayton} et~al.}{1994b}]{1994ApJ...432..785C}
{Clayton} G.~C.,  {Lawson} W.~A.,  {Cottrell} P.~L.,  {Whitney} B.~A.,  {Stanford} S.~A.,   {de Ruyter} F.,  1994b, \mn@doi [\apj] {10.1086/174616}, \href {https://ui.adsabs.harvard.edu/abs/1994ApJ...432..785C} {432, 785}

\bibitem[\protect\citeauthoryear{{Clayton}, {Bjorkman}, {Nordsieck}, {Zellner}  \& {Schulte-Ladbeck}}{{Clayton} et~al.}{1997}]{1997ApJ...476..870C}
{Clayton} G.~C.,  {Bjorkman} K.~S.,  {Nordsieck} K.~H.,  {Zellner} N. E.~B.,   {Schulte-Ladbeck} R.~E.,  1997, \mn@doi [\apj] {10.1086/303666}, \href {https://ui.adsabs.harvard.edu/abs/1997ApJ...476..870C} {476, 870}

\bibitem[\protect\citeauthoryear{{Clayton}, {Geballe}  \& {Zhang}}{{Clayton} et~al.}{2013}]{2013AJ....146...23C}
{Clayton} G.~C.,  {Geballe} T.~R.,   {Zhang} W.,  2013, \mn@doi [\aj] {10.1088/0004-6256/146/2/23}, \href {https://ui.adsabs.harvard.edu/abs/2013AJ....146...23C} {146, 23}

\bibitem[\protect\citeauthoryear{{Cottrell} \& {Lambert}}{{Cottrell} \& {Lambert}}{1982}]{Cottrell1982_rysgrshocks}
{Cottrell} P.~L.,  {Lambert} D.~L.,  1982, The Observatory, \href {https://ui.adsabs.harvard.edu/abs/1982Obs...102..149C} {102, 149}

\bibitem[\protect\citeauthoryear{{Crause}, {Lawson}  \& {Henden}}{{Crause} et~al.}{2007}]{2007MNRAS.375..301C}
{Crause} L.~A.,  {Lawson} W.~A.,   {Henden} A.~A.,  2007, \mn@doi [\mnras] {10.1111/j.1365-2966.2006.11299.x}, \href {https://ui.adsabs.harvard.edu/abs/2007MNRAS.375..301C} {375, 301}

\bibitem[\protect\citeauthoryear{{Crawford} et~al.,}{{Crawford} et~al.}{2023}]{2023MNRAS.521.1674C}
{Crawford} C.~L.,  et~al., 2023, \mn@doi [\mnras] {10.1093/mnras/stad324}, \href {https://ui.adsabs.harvard.edu/abs/2023MNRAS.521.1674C} {521, 1674}

\bibitem[\protect\citeauthoryear{{Crawford}, {Nikultsev}, {Clayton}, {Tisserand}, {Soon}  \& {Pedersen}}{{Crawford} et~al.}{2024}]{Crawford2024_dlhdcmodels}
{Crawford} C.~L.,  {Nikultsev} N.,  {Clayton} G.~C.,  {Tisserand} P.,  {Soon} J.,   {Pedersen} M.~G.,  2024, \mn@doi [\mnras] {10.1093/mnras/stae2149}, \href {https://ui.adsabs.harvard.edu/abs/2024MNRAS.534.1018C} {534, 1018}

\bibitem[\protect\citeauthoryear{{Danziger}}{{Danziger}}{1963}]{Danziger1963_rysgrshocks}
{Danziger} I.~J.,  1963, PhD thesis, -

\bibitem[\protect\citeauthoryear{{De Marco}, {Clayton}, {Herwig}, {Pollacco}, {Clark}  \& {Kilkenny}}{{De Marco} et~al.}{2002}]{2002AJ....123.3387D}
{De Marco} O.,  {Clayton} G.~C.,  {Herwig} F.,  {Pollacco} D.~L.,  {Clark} J.~S.,   {Kilkenny} D.,  2002, \mn@doi [\aj] {10.1086/340569}, \href {https://ui.adsabs.harvard.edu/abs/2002AJ....123.3387D} {123, 3387}

\bibitem[\protect\citeauthoryear{{Feast}}{{Feast}}{1986}]{1986ASSL..128..151F}
{Feast} M.~W.,  1986, IAU Colloq. 87, \href {http://adsabs.harvard.edu/abs/1986ASSL..128..151F} {128, p. 151}

\bibitem[\protect\citeauthoryear{{Feast}}{{Feast}}{1997}]{1997MNRAS.285..339F}
{Feast} M.~W.,  1997, \mn@doi [\mnras] {10.1093/mnras/285.2.339}, \href {https://ui.adsabs.harvard.edu/abs/1997MNRAS.285..339F} {285, 339}

\bibitem[\protect\citeauthoryear{{Feast}, {Carter}, {Roberts}, {Marang}  \& {Catchpole}}{{Feast} et~al.}{1997}]{1997MNRAS.285..317F}
{Feast} M.~W.,  {Carter} B.~S.,  {Roberts} G.,  {Marang} F.,   {Catchpole} R.~M.,  1997, \mn@doi [\mnras] {10.1093/mnras/285.2.317}, \href {https://ui.adsabs.harvard.edu/abs/1997MNRAS.285..317F} {285, 317}

\bibitem[\protect\citeauthoryear{{Feast}, {Whitelock}  \& {Marang}}{{Feast} et~al.}{2003}]{2003MNRAS.346..878F}
{Feast} M.~W.,  {Whitelock} P.~A.,   {Marang} F.,  2003, \mn@doi [\mnras] {10.1111/j.1365-2966.2003.07136.x}, \href {https://ui.adsabs.harvard.edu/abs/2003MNRAS.346..878F} {346, 878}

\bibitem[\protect\citeauthoryear{{Feast}, {Griffin}, {Herbig}  \& {Whitelock}}{{Feast} et~al.}{2019}]{Feast2019_rcrbRVs}
{Feast} M.~W.,  {Griffin} R.~F.,  {Herbig} G.~H.,   {Whitelock} P.~A.,  2019, \mn@doi [\mnras] {10.1093/mnras/sty2893}, \href {https://ui.adsabs.harvard.edu/abs/2019MNRAS.482.4174F} {482, 4174}

\bibitem[\protect\citeauthoryear{{Freytag} \& {H{\"o}fner}}{{Freytag} \& {H{\"o}fner}}{2023}]{Freytag2023_agbdustmodel3D}
{Freytag} B.,  {H{\"o}fner} S.,  2023, \mn@doi [\aap] {10.1051/0004-6361/202244992}, \href {https://ui.adsabs.harvard.edu/abs/2023A&A...669A.155F} {669, A155}

\bibitem[\protect\citeauthoryear{{Gaia Collaboration} et~al.,}{{Gaia Collaboration} et~al.}{2023}]{GaiaDR3_release}
{Gaia Collaboration} et~al., 2023, \mn@doi [\aap] {10.1051/0004-6361/202243940}, \href {https://ui.adsabs.harvard.edu/abs/2023A&A...674A...1G} {674, A1}

\bibitem[\protect\citeauthoryear{{Gail} \& {Sedlmayr}}{{Gail} \& {Sedlmayr}}{1984}]{1984A&A...132..163G}
{Gail} H.~P.,  {Sedlmayr} E.,  1984, \aap, \href {https://ui.adsabs.harvard.edu/abs/1984A&A...132..163G} {132, 163}

\bibitem[\protect\citeauthoryear{{Garc{\'\i}a-Hern{\'a}ndez}, {Rao}, {Lambert}, {Eriksson}, {Reddy}  \& {Masseron}}{{Garc{\'\i}a-Hern{\'a}ndez} et~al.}{2023}]{2023ApJ...948...15G}
{Garc{\'\i}a-Hern{\'a}ndez} D.~A.,  {Rao} N.~K.,  {Lambert} D.~L.,  {Eriksson} K.,  {Reddy} A.~B.~S.,   {Masseron} T.,  2023, \mn@doi [\apj] {10.3847/1538-4357/acc574}, \href {https://ui.adsabs.harvard.edu/abs/2023ApJ...948...15G} {948, 15}

\bibitem[\protect\citeauthoryear{{Goldberg}, {Jiang}  \& {Bildsten}}{{Goldberg} et~al.}{2022}]{2022ApJ...929..156G}
{Goldberg} J.~A.,  {Jiang} Y.-F.,   {Bildsten} L.,  2022, \mn@doi [\apj] {10.3847/1538-4357/ac5ab3}, \href {https://ui.adsabs.harvard.edu/abs/2022ApJ...929..156G} {929, 156}

\bibitem[\protect\citeauthoryear{{Grindlay}, {Tang}, {Los}  \& {Servillat}}{{Grindlay} et~al.}{2012}]{2012IAUS..285...29G}
{Grindlay} J.,  {Tang} S.,  {Los} E.,   {Servillat} M.,  2012, in {Griffin} E.,  {Hanisch} R.,   {Seaman} R.,  eds,  IAU Symposium Vol. 285, New Horizons in Time Domain Astronomy. pp 29--34 (\mn@eprint {arXiv} {1211.1051}), \mn@doi{10.1017/S1743921312000166}

\bibitem[\protect\citeauthoryear{Harris et~al.,}{Harris et~al.}{2020}]{numpy2020}
Harris C.~R.,  et~al., 2020, \mn@doi [Nature] {10.1038/s41586-020-2649-2}, 585, 357

\bibitem[\protect\citeauthoryear{{Hema}, {Pandey}, {Kamath}, {Kameswara Rao}, {Lambert}  \& {Woolf}}{{Hema} et~al.}{2017}]{2017PASP..129j4202H}
{Hema} B.~P.,  {Pandey} G.,  {Kamath} D.,  {Kameswara Rao} N.,  {Lambert} D.,   {Woolf} V.~M.,  2017, \mn@doi [\pasp] {10.1088/1538-3873/aa7f25}, \href {https://ui.adsabs.harvard.edu/abs/2017PASP..129j4202H} {129, 104202}

\bibitem[\protect\citeauthoryear{Hunter}{Hunter}{2007}]{matplotlib2007}
Hunter J.~D.,  2007, Computing in Science \& Engineering, 9, 90

\bibitem[\protect\citeauthoryear{{Ivezi{\'c}} et~al.,}{{Ivezi{\'c}} et~al.}{2019}]{LSST}
{Ivezi{\'c}} {\v{Z}}.,  et~al., 2019, \mn@doi [\apj] {10.3847/1538-4357/ab042c}, \href {https://ui.adsabs.harvard.edu/abs/2019ApJ...873..111I} {873, 111}

\bibitem[\protect\citeauthoryear{{Jayasinghe} et~al.,}{{Jayasinghe} et~al.}{2018}]{2018MNRAS.477.3145J}
{Jayasinghe} T.,  et~al., 2018, \mn@doi [\mnras] {10.1093/mnras/sty838}, \href {https://ui.adsabs.harvard.edu/abs/2018MNRAS.477.3145J} {477, 3145}

\bibitem[\protect\citeauthoryear{{Jayasinghe} et~al.,}{{Jayasinghe} et~al.}{2019}]{2019MNRAS.486.1907J}
{Jayasinghe} T.,  et~al., 2019, \mn@doi [\mnras] {10.1093/mnras/stz844}, \href {https://ui.adsabs.harvard.edu/abs/2019MNRAS.486.1907J} {486, 1907}

\bibitem[\protect\citeauthoryear{{Jurcsik}}{{Jurcsik}}{1996}]{1996AcA....46..325J}
{Jurcsik} J.,  1996, \actaa, \href {https://ui.adsabs.harvard.edu/abs/1996AcA....46..325J} {46, 325}

\bibitem[\protect\citeauthoryear{{Kameswara Rao}, {Ashok}  \& {Kulkarni}}{{Kameswara Rao} et~al.}{1980}]{1980JApA....1...71K}
{Kameswara Rao} I.,  {Ashok} N.~M.,   {Kulkarni} P.~V.,  1980, \mn@doi [Journal of Astrophysics and Astronomy] {10.1007/BF02727951}, \href {https://ui.adsabs.harvard.edu/abs/1980JApA....1...71K} {1, 71}

\bibitem[\protect\citeauthoryear{{Kameswara Rao}, {Lambert}, {Woolf}  \& {Hema}}{{Kameswara Rao} et~al.}{2014}]{2014PASP..126..813K}
{Kameswara Rao} N.,  {Lambert} D.~L.,  {Woolf} V.~M.,   {Hema} B.~P.,  2014, \mn@doi [\pasp] {10.1086/678129}, \href {https://ui.adsabs.harvard.edu/abs/2014PASP..126..813K} {126, 813}

\bibitem[\protect\citeauthoryear{{Karambelkar} et~al.,}{{Karambelkar} et~al.}{2021}]{2021ApJ...910..132K}
{Karambelkar} V.~R.,  et~al., 2021, \mn@doi [\apj] {10.3847/1538-4357/abe5aa}, \href {https://ui.adsabs.harvard.edu/abs/2021ApJ...910..132K} {910, 132}

\bibitem[\protect\citeauthoryear{{Karambelkar}, {Kasliwal}, {Tisserand}, {Clayton}, {Crawford}, {Anand}, {Geballe}  \& {Montiel}}{{Karambelkar} et~al.}{2022}]{2022A&A...667A..84K}
{Karambelkar} V.,  {Kasliwal} M.~M.,  {Tisserand} P.,  {Clayton} G.~C.,  {Crawford} C.~L.,  {Anand} S.~G.,  {Geballe} T.~R.,   {Montiel} E.,  2022, \mn@doi [\aap] {10.1051/0004-6361/202142918}, \href {https://ui.adsabs.harvard.edu/abs/2022A&A...667A..84K} {667, A84}

\bibitem[\protect\citeauthoryear{{Karambelkar} et~al.,}{{Karambelkar} et~al.}{2024}]{2024PASP..136h4201K}
{Karambelkar} V.~R.,  et~al., 2024, \mn@doi [\pasp] {10.1088/1538-3873/ad6210}, \href {https://ui.adsabs.harvard.edu/abs/2024PASP..136h4201K} {136, 084201}

\bibitem[\protect\citeauthoryear{{Kim} et~al.,}{{Kim} et~al.}{2016}]{Kim2016_kmtnet2}
{Kim} S.-L.,  et~al., 2016, \mn@doi [Journal of Korean Astronomical Society] {10.5303/JKAS.2016.49.1.37}, \href {https://ui.adsabs.harvard.edu/abs/2016JKAS...49...37K} {49, 37}

\bibitem[\protect\citeauthoryear{{Kipper} \& {Klochkova}}{{Kipper} \& {Klochkova}}{2005}]{2005BaltA..14..215K}
{Kipper} T.,  {Klochkova} V.~G.,  2005, Baltic Astronomy, \href {https://ui.adsabs.harvard.edu/abs/2005BaltA..14..215K} {14, 215}

\bibitem[\protect\citeauthoryear{{Kipper} \& {Klochkova}}{{Kipper} \& {Klochkova}}{2008}]{2008ASPC..391...43K}
{Kipper} T.,  {Klochkova} V.~G.,  2008, in {Werner} A.,  {Rauch} T.,  eds,  Astronomical Society of the Pacific Conference Series Vol. 391, Hydrogen-Deficient Stars. p.~43

\bibitem[\protect\citeauthoryear{{Kiss}, {Szab{\'o}}  \& {Bedding}}{{Kiss} et~al.}{2006}]{2006MNRAS.372.1721K}
{Kiss} L.~L.,  {Szab{\'o}} G.~M.,   {Bedding} T.~R.,  2006, \mn@doi [\mnras] {10.1111/j.1365-2966.2006.10973.x}, \href {https://ui.adsabs.harvard.edu/abs/2006MNRAS.372.1721K} {372, 1721}

\bibitem[\protect\citeauthoryear{{Kloppenborg}}{{Kloppenborg}}{2023}]{Kloppenborg23}
{Kloppenborg} B.,  2023, https://www.aavso.org

\bibitem[\protect\citeauthoryear{{Lawson} \& {Cottrell}}{{Lawson} \& {Cottrell}}{1997}]{1997MNRAS.285..266L}
{Lawson} W.~A.,  {Cottrell} P.~L.,  1997, \mn@doi [\mnras] {10.1093/mnras/285.2.266}, \href {https://ui.adsabs.harvard.edu/abs/1997MNRAS.285..266L} {285, 266}

\bibitem[\protect\citeauthoryear{{Lawson}, {Cottrell}  \& {Clark}}{{Lawson} et~al.}{1991}]{Lawson1991_rysgrshocks}
{Lawson} W.~A.,  {Cottrell} P.~L.,   {Clark} M.,  1991, \mn@doi [\mnras] {10.1093/mnras/251.4.687}, \href {https://ui.adsabs.harvard.edu/abs/1991MNRAS.251..687L} {251, 687}

\bibitem[\protect\citeauthoryear{{Lawson} et~al.,}{{Lawson} et~al.}{1999}]{1999AJ....117.3007L}
{Lawson} W.~A.,  et~al., 1999, \mn@doi [\aj] {10.1086/300897}, \href {https://ui.adsabs.harvard.edu/abs/1999AJ....117.3007L} {117, 3007}

\bibitem[\protect\citeauthoryear{{Lee}, {Kim}, {Cha}, {Lee}, {Kim}  \& {Park}}{{Lee} et~al.}{2014}]{Lee2014_kmtnet1}
{Lee} C.-U.,  {Kim} S.-L.,  {Cha} S.-M.,  {Lee} Y.,  {Kim} D.-J.,   {Park} B.-G.,  2014, in {Stepp} L.~M.,  {Gilmozzi} R.,   {Hall} H.~J.,  eds,  Society of Photo-Optical Instrumentation Engineers (SPIE) Conference Series Vol. 9145, Ground-based and Airborne Telescopes V. p. 91453T, \mn@doi{10.1117/12.2055571}

\bibitem[\protect\citeauthoryear{{Loreta}}{{Loreta}}{1935}]{1935AN....254..151L}
{Loreta} E.,  1935, Astronomische Nachrichten, \href {http://adsabs.harvard.edu/abs/1935AN....254..151L} {254, 151}

\bibitem[\protect\citeauthoryear{{Marsh}}{{Marsh}}{1976}]{1976ApJ...203..552M}
{Marsh} K.~A.,  1976, \mn@doi [\apj] {10.1086/154111}, \href {https://ui.adsabs.harvard.edu/abs/1976ApJ...203..552M} {203, 552}

\bibitem[\protect\citeauthoryear{Masci et~al.,}{Masci et~al.}{2018}]{Masci2019_ztf}
Masci F.~J.,  et~al., 2018, \mn@doi [Publications of the Astronomical Society of the Pacific] {10.1088/1538-3873/aae8ac}, 131, 018003

\bibitem[\protect\citeauthoryear{{McNaught}}{{McNaught}}{1986}]{1986IAUC.4245....2M}
{McNaught} R.,  1986, \iaucirc, \href {https://ui.adsabs.harvard.edu/abs/1986IAUC.4245....2M} {4245, 2}

\bibitem[\protect\citeauthoryear{{McNaught} \& {Dawes}}{{McNaught} \& {Dawes}}{1986}]{1986IAUC.4233....3M}
{McNaught} R.~H.,  {Dawes} G.,  1986, \iaucirc, \href {https://ui.adsabs.harvard.edu/abs/1986IAUC.4233....3M} {4233, 3}

\bibitem[\protect\citeauthoryear{{Montiel}, {Clayton}, {Sugerman}, {Evans}, {Garcia-Hern{\'a}ndez}, {Kameswara Rao}, {Matsuura}  \& {Tisserand}}{{Montiel} et~al.}{2018}]{2018AJ....156..148M}
{Montiel} E.~J.,  {Clayton} G.~C.,  {Sugerman} B.~E.~K.,  {Evans} A.,  {Garcia-Hern{\'a}ndez} D.~A.,  {Kameswara Rao} N.,  {Matsuura} M.,   {Tisserand} P.,  2018, \mn@doi [\aj] {10.3847/1538-3881/aad772}, \href {https://ui.adsabs.harvard.edu/abs/2018AJ....156..148M} {156, 148}

\bibitem[\protect\citeauthoryear{{O'Keefe}}{{O'Keefe}}{1939}]{1939ApJ....90..294O}
{O'Keefe} J.~A.,  1939, \mn@doi [\apj] {10.1086/144107}, \href {http://adsabs.harvard.edu/abs/1939ApJ....90..294O} {90, 294}

\bibitem[\protect\citeauthoryear{{Payne}}{{Payne}}{1928}]{1928BHarO.861...11P}
{Payne} C.~H.,  1928, Harvard College Observatory Bulletin, \href {https://ui.adsabs.harvard.edu/abs/1928BHarO.861...11P} {861, 11}

\bibitem[\protect\citeauthoryear{{Percy} \& {Dembski}}{{Percy} \& {Dembski}}{2018}]{2018JAVSO..46..127P}
{Percy} J.~R.,  {Dembski} K.~H.,  2018, \mn@doi [\jaavso] {10.48550/arXiv.1809.04484}, \href {https://ui.adsabs.harvard.edu/abs/2018JAVSO..46..127P} {46, 127}

\bibitem[\protect\citeauthoryear{{Percy}, {Bandara}, {Fernie}, {Cottrell}  \& {Skuljan}}{{Percy} et~al.}{2004}]{2004JAVSO..33...27P}
{Percy} J.~R.,  {Bandara} K.,  {Fernie} J.~D.,  {Cottrell} P.~L.,   {Skuljan} L.,  2004, \jaavso, \href {https://ui.adsabs.harvard.edu/abs/2004JAVSO..33...27P} {33, 27}

\bibitem[\protect\citeauthoryear{{Pigott} \& {Englefield}}{{Pigott} \& {Englefield}}{1797}]{1797RSPT...87..133P}
{Pigott} E.,  {Englefield} H.~C.,  1797, Royal Society of London Philosophical Transactions Series I, 87, 133

\bibitem[\protect\citeauthoryear{{Pojmanski}}{{Pojmanski}}{1997}]{Pojmanski1997_ASAS}
{Pojmanski} G.,  1997, \mn@doi [\actaa] {10.48550/arXiv.astro-ph/9712146}, \href {https://ui.adsabs.harvard.edu/abs/1997AcA....47..467P} {47, 467}

\bibitem[\protect\citeauthoryear{{Pugach}}{{Pugach}}{1977}]{1977IBVS.1277....1P}
{Pugach} A.~F.,  1977, Information Bulletin on Variable Stars, \href {https://ui.adsabs.harvard.edu/abs/1977IBVS.1277....1P} {1277, 1}

\bibitem[\protect\citeauthoryear{{Rao} \& {Lambert}}{{Rao} \& {Lambert}}{2003}]{2003PASP..115.1304R}
{Rao} N.~K.,  {Lambert} D.~L.,  2003, \mn@doi [\pasp] {10.1086/379205}, \href {https://ui.adsabs.harvard.edu/abs/2003PASP..115.1304R} {115, 1304}

\bibitem[\protect\citeauthoryear{{Rauer} et~al.,}{{Rauer} et~al.}{2014}]{Plato1}
{Rauer} H.,  et~al., 2014, \mn@doi [Experimental Astronomy] {10.1007/s10686-014-9383-4}, \href {https://ui.adsabs.harvard.edu/abs/2014ExA....38..249R} {38, 249}

\bibitem[\protect\citeauthoryear{{Rauer} et~al.,}{{Rauer} et~al.}{2024}]{Plato2}
{Rauer} H.,  et~al., 2024, \mn@doi [arXiv e-prints] {10.48550/arXiv.2406.05447}, \href {https://ui.adsabs.harvard.edu/abs/2024arXiv240605447R} {p. arXiv:2406.05447}

\bibitem[\protect\citeauthoryear{{Ruiter}, {Ferrario}, {Belczynski}, {Seitenzahl}, {Crocker}  \& {Karakas}}{{Ruiter} et~al.}{2019}]{Ruiter2019}
{Ruiter} A.~J.,  {Ferrario} L.,  {Belczynski} K.,  {Seitenzahl} I.~R.,  {Crocker} R.~M.,   {Karakas} A.~I.,  2019, \mn@doi [\mnras] {10.1093/mnras/stz001}, \href {https://ui.adsabs.harvard.edu/abs/2019MNRAS.484..698R} {484, 698}

\bibitem[\protect\citeauthoryear{{Schaefer}}{{Schaefer}}{2016}]{2016MNRAS.460.1233S}
{Schaefer} B.~E.,  2016, \mn@doi [\mnras] {10.1093/mnras/stw1065}, \href {https://ui.adsabs.harvard.edu/abs/2016MNRAS.460.1233S} {460, 1233}

\bibitem[\protect\citeauthoryear{{Schaefer}}{{Schaefer}}{2024}]{2024MNRAS.527.9274S}
{Schaefer} B.~E.,  2024, \mn@doi [\mnras] {10.1093/mnras/stad3760}, \href {https://ui.adsabs.harvard.edu/abs/2024MNRAS.527.9274S} {527, 9274}

\bibitem[\protect\citeauthoryear{{Schwarzschild}}{{Schwarzschild}}{1975}]{1975ApJ...195..137S}
{Schwarzschild} M.,  1975, \mn@doi [\apj] {10.1086/153313}, \href {https://ui.adsabs.harvard.edu/abs/1975ApJ...195..137S} {195, 137}

\bibitem[\protect\citeauthoryear{{Tisserand}}{{Tisserand}}{2012}]{2012A&A...539A..51T}
{Tisserand} P.,  2012, \mn@doi [\aap] {10.1051/0004-6361/201117874}, \href {https://ui.adsabs.harvard.edu/abs/2012A&A...539A..51T} {539, A51}

\bibitem[\protect\citeauthoryear{{Tisserand} et~al.,}{{Tisserand} et~al.}{2009}]{2009A&A...501..985T}
{Tisserand} P.,  et~al., 2009, \mn@doi [\aap] {10.1051/0004-6361/200911808}, \href {https://ui.adsabs.harvard.edu/abs/2009A&A...501..985T} {501, 985}

\bibitem[\protect\citeauthoryear{{Tisserand}, {Clayton}, {Welch}, {Pilecki}, {Wyrzykowski}  \& {Kilkenny}}{{Tisserand} et~al.}{2013}]{2013A&A...551A..77T}
{Tisserand} P.,  {Clayton} G.~C.,  {Welch} D.~L.,  {Pilecki} B.,  {Wyrzykowski} L.,   {Kilkenny} D.,  2013, \mn@doi [\aap] {10.1051/0004-6361/201220713}, \href {https://ui.adsabs.harvard.edu/abs/2013A&A...551A..77T} {551, A77}

\bibitem[\protect\citeauthoryear{{Tisserand} et~al.,}{{Tisserand} et~al.}{2020}]{2020A&A...635A..14T}
{Tisserand} P.,  et~al., 2020, \mn@doi [\aap] {10.1051/0004-6361/201834410}, \href {https://ui.adsabs.harvard.edu/abs/2020A&A...635A..14T} {635, A14}

\bibitem[\protect\citeauthoryear{{Tisserand} et~al.,}{{Tisserand} et~al.}{2022}]{2022A&A...667A..83T}
{Tisserand} P.,  et~al., 2022, \mn@doi [\aap] {10.1051/0004-6361/202142916}, \href {https://ui.adsabs.harvard.edu/abs/2022A&A...667A..83T} {667, A83}

\bibitem[\protect\citeauthoryear{{Tisserand}, {Crawford}, {Soon}, {Clayton}, {Ruiter}  \& {Seitenzahl}}{{Tisserand} et~al.}{2024a}]{2024A&A...684A.130T}
{Tisserand} P.,  {Crawford} C.~L.,  {Soon} J.,  {Clayton} G.~C.,  {Ruiter} A.~J.,   {Seitenzahl} I.~R.,  2024a, \mn@doi [\aap] {10.1051/0004-6361/202348004}, \href {https://ui.adsabs.harvard.edu/abs/2024A&A...684A.130T} {684, A130}

\bibitem[\protect\citeauthoryear{{Tisserand}, {Crawford}, {Soon}, {Clayton}, {Ruiter}  \& {Seitenzahl}}{{Tisserand} et~al.}{2024b}]{Tisserand2024_3Ddistribution}
{Tisserand} P.,  {Crawford} C.~L.,  {Soon} J.,  {Clayton} G.~C.,  {Ruiter} A.~J.,   {Seitenzahl} I.~R.,  2024b, \mn@doi [\aap] {10.1051/0004-6361/202348005}, \href {https://ui.adsabs.harvard.edu/abs/2024A&A...684A.131T} {684, A131}

\bibitem[\protect\citeauthoryear{{Tonry} et~al.,}{{Tonry} et~al.}{2018}]{Tonry2018_atlas}
{Tonry} J.~L.,  et~al., 2018, \mn@doi [\pasp] {10.1088/1538-3873/aabadf}, \href {https://ui.adsabs.harvard.edu/abs/2018PASP..130f4505T} {130, 064505}

\bibitem[\protect\citeauthoryear{{Udalski}}{{Udalski}}{2008}]{2008AcA....58..187U}
{Udalski} A.,  2008, \mn@doi [\actaa] {10.48550/arXiv.0810.2244}, \href {https://ui.adsabs.harvard.edu/abs/2008AcA....58..187U} {58, 187}

\bibitem[\protect\citeauthoryear{Virtanen et~al.,}{Virtanen et~al.}{2020}]{scipy2020}
Virtanen P.,  et~al., 2020, \mn@doi [Nature Methods] {10.1038/s41592-019-0686-2}, \href {https://rdcu.be/b08Wh} {17, 261}

\bibitem[\protect\citeauthoryear{{Wenzel}, {Hoffmeister}  \& {McNaught}}{{Wenzel} et~al.}{1986}]{1986IAUC.4241....2W}
{Wenzel} W.,  {Hoffmeister} C.,   {McNaught} R.~H.,  1986, \iaucirc, \href {https://ui.adsabs.harvard.edu/abs/1986IAUC.4241....2W} {4241, 2}

\bibitem[\protect\citeauthoryear{{Wheatland}}{{Wheatland}}{2000}]{Wheatland2000_solarflareoccurrence}
{Wheatland} M.~S.,  2000, \mn@doi [\apjl] {10.1086/312739}, \href {https://ui.adsabs.harvard.edu/abs/2000ApJ...536L.109W} {536, L109}

\bibitem[\protect\citeauthoryear{{Whitelock}, {Feast}, {Marang}  \& {Overbeek}}{{Whitelock} et~al.}{1997}]{1997MNRAS.288..512W}
{Whitelock} P.~A.,  {Feast} M.~W.,  {Marang} F.,   {Overbeek} M.~D.,  1997, \mn@doi [\mnras] {10.1093/mnras/288.2.512}, \href {https://ui.adsabs.harvard.edu/abs/1997MNRAS.288..512W} {288, 512}

\bibitem[\protect\citeauthoryear{{Whitelock}, {Feast}, {Marang}  \& {Groenewegen}}{{Whitelock} et~al.}{2006}]{2006MNRAS.369..751W}
{Whitelock} P.~A.,  {Feast} M.~W.,  {Marang} F.,   {Groenewegen} M.~A.~T.,  2006, \mn@doi [\mnras] {10.1111/j.1365-2966.2006.10322.x}, \href {https://ui.adsabs.harvard.edu/abs/2006MNRAS.369..751W} {369, 751}

\bibitem[\protect\citeauthoryear{{Whitney}, {Clayton}, {Schulte-Ladbeck}  \& {Meade}}{{Whitney} et~al.}{1992}]{1992AJ....103.1652W}
{Whitney} B.~A.,  {Clayton} G.~C.,  {Schulte-Ladbeck} R.~E.,   {Meade} M.~R.,  1992, \mn@doi [\aj] {10.1086/116180}, \href {https://ui.adsabs.harvard.edu/abs/1992AJ....103.1652W} {103, 1652}

\bibitem[\protect\citeauthoryear{{Whitney}, {Balm}  \& {Clayton}}{{Whitney} et~al.}{1993}]{1993ASPC...45..115W}
{Whitney} B.~A.,  {Balm} S.~P.,   {Clayton} G.~C.,  1993, in {Sasselov} D.~D.,  ed.,  Astronomical Society of the Pacific Conference Series Vol. 45, Luminous High-Latitude Stars. p.~115

\bibitem[\protect\citeauthoryear{{Woitke}, {Goeres}  \& {Sedlmayr}}{{Woitke} et~al.}{1996}]{1996A&A...313..217W}
{Woitke} P.,  {Goeres} A.,   {Sedlmayr} E.,  1996, \aap, \href {https://ui.adsabs.harvard.edu/abs/1996A&A...313..217W} {313, 217}

\bibitem[\protect\citeauthoryear{{Yakovina}, {Pugach}  \& {Pavlenko}}{{Yakovina} et~al.}{2009}]{Yakovina2009_dyperabunds}
{Yakovina} L.~A.,  {Pugach} A.~F.,   {Pavlenko} Y.~V.,  2009, \mn@doi [Astronomy Reports] {10.1134/S1063772909030019}, \href {https://ui.adsabs.harvard.edu/abs/2009ARep...53..187Y} {53, 187}

\bibitem[\protect\citeauthoryear{{Yuin}}{{Yuin}}{1948}]{1948ApJ...107..413Y}
{Yuin} C.,  1948, \mn@doi [\apj] {10.1086/145028}, \href {https://ui.adsabs.harvard.edu/abs/1948ApJ...107..413Y} {107, 413}

\bibitem[\protect\citeauthoryear{{Za{\v{c}}s}, {Mondal}, {Chen}, {Pugach}, {Musaev}  \& {Alksnis}}{{Za{\v{c}}s} et~al.}{2007}]{Zacs2007_dypersolar}
{Za{\v{c}}s} L.,  {Mondal} S.,  {Chen} W.~P.,  {Pugach} A.~F.,  {Musaev} F.~A.,   {Alksnis} O.,  2007, \mn@doi [\aap] {10.1051/0004-6361:20066923}, \href {https://ui.adsabs.harvard.edu/abs/2007A&A...472..247Z} {472, 247}

\makeatother
\end{thebibliography}

\bsp	
\label{lastpage}
\end{document}